\documentclass[11pt]{article} 
\pdfoutput=1

\usepackage{amsmath}
\usepackage[norelsize]{algorithm2e}
\usepackage{scalerel}
\usepackage{tikz-cd}
\usepackage{ulem}
\usepackage{float}
\restylefloat{figure}
\usepackage{stackengine}
\usepackage[english]{babel}
\usepackage[light,condensed,math]{anttor}
\usepackage[T1]{fontenc}
\usepackage{kpfonts}
\usepackage{comment}
\usepackage{array}
\usepackage{lmodern}
\usepackage[mathscr]{euscript}

\usepackage{breqn}
\usepackage{todonotes}
\usepackage{slashed}
\usepackage{youngtab}
\usepackage{amsmath}
\usepackage{amssymb}
\usepackage{amsxtra}
\usepackage{color}
\usepackage[margin=1.2in]{geometry}

\newcommand {\Det} {\tt Det}
\newcommand{\mA}{{\mathscr A}}

\newcommand {\bla} {\underline{\boldsymbol{\lambda}}}

\newcommand{\ve}{{\varepsilon}}
\renewcommand{\Im}{{\operatorname{Im}\,}}

\newcommand{\BC}{{\mathbb{C}}}

\newcommand{\BP}{{\mathbb{P}}}
\newcommand{\BQ}{{\mathbb{Q}}}

\newcommand {\mt} {\tt m}
\newcommand{\y}{{\mathscr Y}}
\newcommand{\mV}{{\mathscr V}}
\newcommand{\x}{{\mathscr X}}
\newcommand{\bfla}{\mathbf{\Lambda}}

\newcommand{\om}{\omega}
\newcommand{\my}[1]{{\mathscr Y} \left( #1 \right)}

\newcommand {\Nq} {\mathscr\nabla^{\qe}}

\newcommand {\wt} {\tt w}

\newcommand {\tw} {\text{w}}
\newcommand{\Bfn}{{\CalB}\, \left( \begin{matrix} {\ve}_{3} &   {\ve}_{4} \\
 {\ve}_{1} & {\ve}_{2} \end{matrix} \, \biggl\vert \,  \begin{matrix}  {\mathfrak s} \\
  p \end{matrix} \, \biggr\vert\  {\bqt}\  \right)}
 \newcommand{\Qfn}{{\CalQ}\, \left( \begin{matrix} {\omega} & {\mathfrak s} \\
 & p \end{matrix} \, \biggr\vert\  {\bqt}\  \right)}
  \newcommand{\tQfn}{{\tilde\CalQ}\, \left( \begin{matrix} {\omega} & {\mathfrak s} \\
 & p \end{matrix} \, \biggr\vert\  {\bqt}\  \right)}
\newcommand {\bfc} {\underline{\fc}}

\newcommand {\bmt} {\underline{\boldsymbol{\mt}}}

\newcommand {\ept} {\underline{\ve}}
\newcommand {\rv} {\underline{\rho}}
\newcommand {\bzv} {\underline{\mathbf{z}}}

\newcommand {\bqt} {\underline{\qe}}
\newcommand {\Dq} {\mathscr D^{\qe}}

\newcommand {\bkt} {\underline{\bk}}

\newcommand{\sdtimes}{\mathbin{
\hbox{\hskip2pt
\vrule height 4.1pt depth -.3pt width.25pt\hskip-2pt$\times$}}}

\newcommand{\vev}[1]{\left\langle\ #1 \ \right\rangle}

\newcommand{\Gg}{\mathsf{G}_{\mathbf{g}}}

\newcommand {\ac} {\mathfrak{a}}

\newcommand {\fb} {\mathfrak{b}}
\newcommand {\fc} {\mathfrak{c}}

\newcommand {\fe} {\mathfrak{f}}

\def\beq{\begin{equation}}
\def\eeq{\end{equation}}

\newcommand{\ba} {\mathbf{a}}
\newcommand {\BB}   {\mathbb B}

\newcommand {\bQ}   {\mathbf{Q}}

\newcommand {\qe} {\mathfrak q}

\newcommand {\ib} {\mathbf{i}}

\newcommand {\bnu}{ {\boldsymbol{\nu}}}

\newcommand {\bk}{ \mathbf{k}}

\newcommand {\bz}{ \mathbf{z}}

\newcommand {\bS}{ \mathbf{S}}

\newcommand {\BDe}   {\boldsymbol{\Delta}}

\newcommand {\BZ}   {\mathbb Z}

\newcommand {\CalB} {\mathcal B}

\newcommand {\CalD} {\mathcal D}
\newcommand {\CalE} {\mathcal E}

\newcommand {\CalK} {\mathcal K}

\newcommand {\CalM} {\mathcal M}
\newcommand {\CalN} {\mathcal N}
\newcommand {\CalO} {\mathcal O}

\newcommand {\CalQ} {\mathcal Q}
\newcommand {\CalR} {\mathcal R}

\newcommand {\CalX} {\mathcal X}

\newcommand {\CalW} {\mathcal W}
\newcommand {\CalZ} {\mathcal Z}

\newcommand{\fo}{\vert\kern -.03in\_}

\newcommand {\ii} {\mathrm{i}}

\newcommand {\4}{\underline{\bf 4}}
\newcommand {\6}{\underline{\bf 6}}
\usepackage{float}
\restylefloat{figure}

\newcommand{\boxit}[1]{\vbox{\hrule\hbox{\vrule\kern8pt
\vbox{\hbox{\kern8pt}\hbox{\vbox{#1}}\hbox{\kern8pt}}
\kern8pt\vrule}\hrule}}
\newcommand{\mathboxit}[1]{\vbox{\hrule\hbox{\vrule\kern8pt\vbox{\kern8pt
\hbox{$\displaystyle #1$}\kern8pt}\kern8pt\vrule}\hrule}}

\newcommand{\picti}[2]{\includegraphics[width=#1cm]{#2.pdf}}

\title{\textsc{BPS/CFT correspondence V: \\ BPZ and KZ
equations from $qq$-characters}}
\author{Nikita Nekrasov\footnote{Simons Center for Geometry and
    Physics, Stony Brook University, Stony Brook, NY 11794\, , 
    \newline{\tiny on leave of absence from:} 
    Kharkevich IITP RAS, Moscow, ITEP, Moscow .\newline
    e-mail: nnekrasov@scgp.stonybrook.edu}}
\date{}

\begin{document}
\maketitle

\begin{abstract}
 
We illustrate the use of the theory of $qq$-characters by deriving the BPZ and KZ-type equations for the partition functions of certain surface defects in quiver ${\CalN}=2$ theories. We generate a surface defect
in the linear quiver theory by embedding it into a theory with additional node, with specific masses of the fundamental hypermultiplets at that node. We prove
that the supersymmetric partition function of this theory with $SU(2)^{r-3}$ gauge group verifies the celebrated Belavin-Polyakov-Zamolodchikov equation
of two dimensional Liouville theory. We also study the $SU(n)$ theory with $2n$ fundamental hypermultiplets and the theory with adjoint hypermultiplet. We show that the regular orbifold defect in this theory solves the KZ-like equation of the WZW theory on a four punctured sphere and one-punctured torus, respectively. In the companion paper \cite{NT} these equations will be mapped to the Knizhnik-Zamolodchikov equations

\end{abstract}

\section{Introduction}

In the recent paper \cite{N3} we attempted to study the
 novel type of Dyson-Schwinger identities, relating non-perturbative contributions to the correlation functions coming from different topological sectors of the field space. Our main tool is the class of special observables called the $qq$-characters, which are most useful for exploiting these identities. In this paper we shall look at the examples of applications of these identities.

\section{Linear quiver theories}
 
In this section we will be dealing with the linear $A$-type quiver ${\CalN}=2$  gauge theories in four dimensions. The gauge group of the corresponding theory is a product of $r$ factors,
\beq
{\Gg} = U(N) \times \ldots \times U(N)
\label{eq:line}
\eeq
which we shall label by the vertices $i = 1, \ldots , r$ of the $A_{r}$ Dynkin
diagram. The theory has matter hypermultiplet fields, transforming in the bifundamental representations $(N_{i}, {\bar N}_{i+1})$, and two pairs of
$N$ fundamental hypermultiplets, charged under the $U(N)_{1}$ and the
$U(N)_{r}$ gauge factors, respectively. 
The Lagrangian of the theory and the choice of the vacuum is parametrized by the complexified gauge couplings
\beq
{\tau}_{i} = \frac{{\vartheta}_{i}}{2\pi} + \frac{4\pi \ii}{g_{i}^{2}} \in H_{+}, 
\qquad {\qe}_{i} = {\exp} \, 2\pi \ii {\tau}_{i} \in {\BC}^{\times}_{| \cdot | < 1}, \qquad 
\eeq
and
the Coulomb moduli ${\ac}_{i} = ( {\ac}_{i, \alpha} )_{{\alpha}=1}^{N}$, where the range of the index $i$ is extended to $0 \leq i \leq r+1$ with the convention where
${\ac}_{0, \alpha}$ and ${\ac}_{r+1, \alpha}$ encode the masses of the fundamental hypermultiplets, while ${\qe}_{0} = {\qe}_{r+1} = 0$. 

In this paper, as in \cite{N3} we study vacuum expectation values of the composite operators, the $qq$-characters, which are built out of the $\y$-observables. For each node $i$, $i = 0 , \ldots , r+1$
\beq
{\y}_{i}(x) \sim {\rm Det} ( x- {\Phi}_{i} ) 
\label{eq:yix}
\eeq
is the regularized characteristic polynomial of the adjoint complex scalar in the vector multiplet of the $i$'th factor of the gauge group (which is actually 
the flavor group for $i = 0, r+1$). 

The calculations in gauge theory are facilitated by $\Omega$-deformation and the
use of equivariant localization \cite{N2}. The partition
function we are going to study is the sum over the toric instantons, which
are in one-to-one correspondence with the 
collections ${\bfla} = ({\lambda}^{(i, {\alpha})})$, $i = 1, \ldots , r$, ${\alpha} = 1, \ldots , N$ of Young diagrams:
\beq
{\lambda}^{(i,{\alpha})} = \left( {\lambda}^{(i,{\alpha})}_{1} \geq {\lambda}^{(i,{\alpha})}_{2} \geq \ldots \geq {\lambda}^{(i,{\alpha})}_{{\ell}({\lambda}^{(i,{\alpha})})} > 0 \right)
\ . 
\eeq
The $\y$-observables \eqref{eq:yix} evaluate on $\bfla$ to 
\beq
{\y}_{i}(x) \vert_{\bfla} = \prod_{{\alpha}=1}^{N} \left( ( x - {\ac}_{i, \alpha} ) \, \prod_{\square \in {\lambda}^{(i, {\alpha})}} \frac{(x  - c_{\square} - {\ve}_{1})(x  - c_{\square} - {\ve}_{2})}{(x  - c_{\square})(x  - c_{\square} - {\ve})} \right)
\label{eq:yixla}
\eeq
Here ${\ve} = {\ve}_{1} + {\ve}_{2}$, ${\ve}_{1}, {\ve}_{2}$ are the two complex parameters of the $\Omega$-background, 
${\square} \equiv (a,b) \in {\lambda}^{(i,{\alpha})} \, \Leftrightarrow $
\beq
1 \leq b \leq {\lambda}^{(i,{\alpha})}_{a}, \qquad
1 \leq a \leq \left( {\lambda}^{(i,{\alpha})}_{b} \right)^{t}
\eeq
and
\beq
c_{\square} \equiv {\ac}_{i, \alpha} + {\ve}_{1}(a-1) + {\ve}_{2}(b-1)
\label{eq:cont}
\eeq
The supersymmetric partition function of the theory we discuss
is given by the product:
\beq
{\CalZ}_{N} (\mathbf{\ac}, \mathbf{\qe}, {\ve}_{1}, {\ve}_{2} ) = {\CalZ}_{N}  (\mathbf{\ac};  \mathbf{\qe};  {\ve}_{1}, {\ve}_{2} )^{\rm tree} {\CalZ}_{N}  (\mathbf{\ac};  {\ve}_{1}, {\ve}_{2} )^{\rm 1-loop} {\CalZ}_{N}  (\mathbf{\ac}, \mathbf{\qe}; {\ve}_{1}, {\ve}_{2} )^{\rm inst}
\label{eq:czquiv}
\eeq
where
\beq
{\CalZ}_{N}  (\mathbf{\ac};  \mathbf{\qe};  {\ve}_{1}, {\ve}_{2} )^{\rm tree}  = \prod_{i=1}^{r} \prod_{\alpha = 1}^{N} \, {\qe}_{i}^{-\frac{1}{2{\ve}_{1}{\ve}_{2}} {\ac}_{i,\alpha}^{2}}
\label{eq:cztree}
\eeq
\beq
{\CalZ}_{N}  (\mathbf{\ac};  {\ve}_{1}, {\ve}_{2} )^{\rm 1-loop} = 
\prod_{{\alpha},{\beta}=1}^{N} \frac{\prod_{i=1}^{r} {\Gamma}_{2}({\ac}_{i,\alpha} - {\ac}_{i,\beta} ; {\ve}_{1}, {\ve}_{2} )}{\prod_{i=1}^{r+1} {\Gamma}_{2}({\ac}_{i-1,\beta} - {\ac}_{i,\alpha}; {\ve}_{1}, {\ve}_{2} )} \, 
\label{eq:1loop}
\eeq
with ${\Gamma}_{2}(x ; {\ve}_{1}, {\ve}_{2})$ the Barnes double gamma function, an entire function of $x$ with simple zeroes at $x = m {\ve}_{1} + n {\ve}_{2}$, $n, m \in {\BZ}_{> 0}$. Finally, 
\beq
{\CalZ}_{N}  (\mathbf{\ac}, \mathbf{\qe}; {\ve}_{1}, {\ve}_{2} )^{\rm inst}
= \sum_{\bfla} {\CalQ}_{\bfla} \ {\CalM}_{\bfla} \eeq
where we set ${\lambda}^{(0,{\alpha})} = {\lambda}^{(r+1, {\alpha})} = {\emptyset}$, 
\beq
{\CalQ}_{\bfla} =  \prod_{i=1}^{r} \prod_{\alpha = 1}^{N} \ {\qe}_{i}^{|{\lambda}^{(i, {\alpha})}|}
\label{eq:instq}
\eeq
and
\beq
{\CalM}_{\bfla} = 
\prod_{{\alpha},{\beta}=1}^{N} \frac{\prod_{i=1}^{r+1} {\CalM}_{{\ve}_{1}, {\ve}_{2}}({\ac}_{i-1,\alpha} - {\ac}_{i,\beta}; {\lambda}^{(i-1,{\alpha})}, {\lambda}^{(i,{\beta})})}{\prod_{i=1}^{r} {\CalM}_{{\ve}_{1}, {\ve}_{2}}({\ac}_{i,\alpha} - {\ac}_{i,\beta} ; {\lambda}^{(i,{\alpha})}, {\lambda}^{(i,{\beta})} )}
\label{eq:cmbfla}
\eeq
with
\beq
{\CalM}_{{\ve}_{1}, {\ve}_{2}}(x; {\lambda}, {\nu}) = \prod_{(a,b) \in {\nu}} \left( x + {\ve}_{1} ( a - {\nu}^{t}_{b}) + {\ve}_{2} ( {\lambda}_{a}+1- b) \right) \prod_{(a,b) \in {\lambda}}  \left( x + {\ve}_{1} ( {\lambda}^{t}_{b}+1-a) + {\ve}_{2} ( b - {\nu}_{a}) \right)
\eeq
Note that
\beq
{\CalM}_{{\ve}_{1}, {\ve}_{2}}(x; {\lambda}, {\emptyset}) = P_{\lambda}(x + {\ve}), \quad {\CalM}_{{\ve}_{1}, {\ve}_{2}}(x; {\emptyset}, {\nu}) = (-1)^{|{\nu}|}
P_{\nu} ( -x )
\eeq
where
\beq
P_{\lambda}(x) \equiv \prod_{(a,b) \in {\lambda}}  \left( x + {\ve}_{1} ( a-1) + {\ve}_{2} ( b-1) \right) 
\label{eq:content}
\eeq
is the so-called content polynomial. 
\subsection{Non-perturbative Dyson-Schwinger identities}

The main statement of \cite{N3} is that the expectation values 
$\vev{{\CalX}_{i}(x)}_{N} $ of the $qq$-character observables:
\beq
{\CalX}_{i}(x) = {\y}_{0}(x) \, \sum_{J \subset [0,r]} \prod_{j \in J} \, {\Xi}_{j}(x+{\ve}h_{j})
\label{eq:qqchar}
\eeq
are entire functions of $x$, in fact, polynomials of degree $N$ (cf. \eqref{eq:symz}):
\beq
 \vev{{\CalX}_{i}(x)} = T_{i}(x) = e_{i} x^{N} + \sum_{p=1}^{N} t_{i,p}x^{N-p}
 \label{eq:tix}
\eeq
Here
\beq
{\Xi}_{i}(x) = z_{i} \frac{{\y}_{i+1}(x+{\ve})}{{\y}_{i}(x)}\, , \qquad i = 0, \ldots r
\label{eq:xifromyx}
\eeq
the parameters $z_{i}$ are related to $\qe_{i}$ via:
\beq
{\qe}_{i} = z_{i}/z_{i-1}, \qquad i = 0, \ldots, r+1
\label{eq:zfroq}
\eeq
so that $z_{-1} = \infty$, $z_{r+1} = 0$ and $z_{0}, z_{1}, \ldots , z_{r}$
are defined up to  an overall rescaling. Finally, the symbol $\vev{{\CalO}}$, for a function ${\CalO} = {\CalO}_{\bfla}$ on the set of $N\times r$-tuples of partitions  is defined as the complexified statistical average:
\beq
\vev{{\CalO}}_{N}  = \frac{1}{{\CalZ}_{N}  (\mathbf{\ac}, \mathbf{\qe}; {\ve}_{1}, {\ve}_{2} )^{\rm inst}}\, 
 \sum_{\bfla} \ {\CalQ}_{\bfla} \ {\CalM}_{\bfla} \ {\CalO}_{\bfla}
\eeq

\subsection{BPZ equation}

Consider the $r+3$-point conformal block 
\beq {\chi}_{r+3}({\bz}) =  \vev{  \prod_{i=-1}^{r+1} {\mV}_{{\Delta}_{i}}(z_{i})}^{\rm chiral} \label{eq:degcb}
\eeq
of conformal primary operators with dimensions ${\Delta}_{i}$, $i = -1, \ldots , r+1$ in the two dimensional conformal Liouville theory 
with the Virasoro central charge 
\beq
 c = 1 + 6 \left( b + b^{-1} \right)^2 \label{eq:liouvcc}
 \eeq
 When one of the primary dimensions, e.g. ${\Delta}_{0}$
 is equal to either  
\beq
\Delta_{1,2} = -\frac{1}{2}-\frac{3b^2}{4}\,, \qquad {\rm or}\qquad
\Delta_{1,2} = -\frac{1}{2}-\frac{3}{4b^2}\, \ ,\eeq
then the corresponding fields ${\mV}_{\Delta_{0}}$ have null-vectors among their descendants. 
The decoupling of the null-vectors
 implies  the second-order differential equation for the conformal
blocks, the BPZ equation \cite{Belavin:1984vu}: 
\begin{eqnarray}
\left[\,\frac{3}{2 ( 2{\Delta}_{0}+1 )}\,\frac{\partial^2}{\partial z_{0}^2} + \sum_{i \neq 0}
\left(\,\frac{\Delta_i}{(z_{0}-z_i)^2}+\frac{1}{z_{0}-z_i}\,\frac{\partial}{\partial z_{i}}
\,\right)\,\right]\, {\chi}_{r+3} ({\bz}) =0\, .
\label{eq:bpz}
\end{eqnarray}
where $i = -1, 0, 1, \ldots, r+1$. 
In addition, \eqref{eq:degcb} obeys the global Virasoro invariance constraints:
\beq\begin{aligned}
& \left[\, z_{0}^{-1} {\nabla}^{z}_{0}  +  
\sum_{i\neq 0} z_{i}^{-1} {\nabla}^{z}_{i}  \right]\,  {\chi}_{r+3} ({\bz})   = 0 \\
&  \left[\, {\nabla}^{z}_{0} +  {\Delta}_{0}   +  \sum_{i \neq 0} \left( {\nabla}^{z}_{i}  + {\Delta}_{i} \right) \right]\,  {\chi}_{r+3} ({\bz})   = 0  \\
& \left[\, z_{0} \left( {\nabla}^{z}_{0} +  2 {\Delta}_{0} \right)   +  \sum_{i \neq 0} z_i \left( {\nabla}^{z}_i  + 2 {\Delta}_i \right) \right]\,
{\chi}_{r+3} ({\bz})   = 0 \ . 
\label{eq:vir3}\end{aligned}\eeq
where
\beq
{\nabla}_{i}^{z} = z_{i} \frac{\partial}{{\partial}z_{i}}
\label{eq:nabi}
\eeq
Let us express ${\nabla}^{z}_{-1} {\chi}$ and ${\nabla}^{z}_{r+1} {\chi}$ in terms of ${\nabla}_{0}^{z}{\chi}$ and ${\nabla}_{i}^{z}{\chi}$, $i = 1, \ldots , r$
using the first and the third equations in \eqref{eq:vir3}. 
The equation \eqref{eq:bpz} and the second equation in \eqref{eq:vir3}
now become the partial differential equations in $r+1$ variables $z_{0}, z_{1}, \ldots , z_{r}$, with
$z_{-1}, z_{r+1}$ playing the role of parameters. 
Now set 
\beq
b^{2} = {\ve}_{2}/{\ve}_{1}
\eeq
and
\beq
{\Delta}_{0} =  - \frac{1}{2} - \frac{3{\ve}_{2}}{4 {\ve}_{1}} 
\label{eq:eed0}
\eeq
Without any loss of generality we can now set $z_{-1} = {\infty}$ and $z_{r+1} = 0$ in \eqref{eq:bpz}, with the result:
\beq
\begin{aligned}
& \left[  \left( {\ve}_{1} {\nabla}_{0}^{z} \right)   \left( {\ve}_{1}{\nabla}_{0}^{z}  - {\ve} \right) +
{\ve}_{1}{\ve}_{2} \left( {\Delta}_{r+1} + \sum_{i=1}^{r} \left( {\Delta}_{i} u_{i}^{2}  -  u_{i} {\nabla}_{i}^{z} \right)  \right) \right] \, {\chi}_{r+3} = 0 \\
& \left[ {\nabla}_{0}^{z} + {\Delta}_{0} - {\Delta}_{-1} + {\Delta}_{r+1} + \sum_{i=1}^{r} \left( {\nabla}_{i}^{z} + {\Delta}_{i} \right) \right] \, {\chi}_{r+3} = 0 \\
\end{aligned}
\label{eq:bpz2}
\eeq
where
\beq
u_{i} = \frac{z_{0}}{z_{i} - z_{0}}\, , \quad i = 1, \ldots , r \ .
\label{eq:ui}
\eeq
In this paper we prove that 
 the properly normalized partition function
 \beq
 {\CalZ}_{2}/{\CalZ}_{1}
 \eeq
 (we fix the normalization of the parameters below)
 solves the BPZ equation  \eqref{eq:bpz2} when the parameters $\mathbf{{\ac}_{0}}$ and $\mathbf{{\ac}_{1}}$ are in the following relation:
 For some $\alpha = 1, \ldots , N$
 \beq
 \begin{aligned}
&  {\ac}_{0,{\beta}} = {\ac}_{1, \beta}, \qquad {\beta} \neq \alpha
 \\
 & {\ac}_{0, \alpha} = {\ac}_{1, \alpha} + {\ve}_{2} 
 \\
 \end{aligned}
 \label{eq:dege}
 \eeq

\subsection{Remarks}
\begin{enumerate}

\item

In this paper we study the case $N=2$. The $N>2$ generalization is possible,
the Liouville theory being replaced by the $A_{N-1}$ Toda theory. 

\item

The condition \eqref{eq:dege} depends on the choice of $\alpha$. There
are, therefore, $N$ solutions of \eqref{eq:bpz2}, corresponding to the $N$ choices of $\alpha$. 

\item

The degeneration \eqref{eq:dege} is not unique. In fact, \eqref{eq:dege} can be generalized to:
For some $\alpha = 1, \ldots , N$ and $i = 1, \ldots, r+1$
 \beq
 \begin{aligned}
&  {\ac}_{i-1,{\beta}} = {\ac}_{i, \beta}, \qquad {\beta} \neq \alpha
 \\
 & {\ac}_{i-1, \alpha} = {\ac}_{i, \alpha} + {\ve}_{2} 
 \\
 \end{aligned}
 \label{eq:dege2}
 \eeq
 We will not show in this paper that the partition functions which are obtained by the tuning \eqref{eq:dege2} of the Coulomb moduli provide the local solutions
 to \eqref{eq:bpz2} with $z_{0} \leftrightarrow z_{i}$, ${\Delta}_{0} \leftrightarrow
 {\Delta}_{i}$, which are analytic in the domain $|z_{i-2}| > |z_{i-1}| > |z_{i}|$. 

\item

The relation \eqref{eq:dege} can be also generalized to
 \beq
 \begin{aligned}
&  {\ac}_{0,{\beta}} = {\ac}_{1, \beta}, \qquad {\beta} \neq \alpha
 \\
 & {\ac}_{0, \alpha} = {\ac}_{1, \alpha} + (p-1){\ve}_{2} + (q-1) {\ve}_{1}
 \\
 \end{aligned}
 \label{eq:dege3}
 \eeq
with $p,q > 1$. When $p$ and $q$ are mutually prime the equation \eqref{eq:bpz} is generalized to the differential
equation of the order $pq$. 

\end{enumerate}

\section{Derivation}

The strategy of our derivation is simple. Expand ${\y}_{i}(x)$, ${\Xi}_{i}(x)$ and ${\CalX}_{i}(x)$ at infinity in $x$. Specifically, we are interested in the $x^{-1}$ coefficient ${\CalX}_{i}^{(-1)}$ of the large $x$ expansion of ${\CalX}_{i}(x)$. The  main theorem \eqref{eq:tix} states that 
\beq
\vev{{\CalX}_{i}^{(-1)}} = 0, \qquad i = 1, \ldots , r
\label{eq:cixeq}
\eeq
It is convenient to package \eqref{eq:cixeq} into the equation:
\beq
\prod_{j=0}^{r} \frac{1}{1+tz_{j}} \ \sum_{i=0}^{r+1} t^{i} \vev{{\CalX}_{i}^{(-1)}} = 0
\label{eq:tcix}
\eeq
 Now, from \eqref{eq:yixla} we compute:
 \begin{multline}
 {\log} {\Xi}_{i}(x) \vert_{\bfla} = {\log} \left( z_{i} \right) +  \\ {\exp} \, \frac{1}{x} A_{i}^{(1)} + \frac{1}{x^{2}} \left( \frac 12
 A_{i}^{(2)} + {\ve}_{1}{\ve}_{2} (k_{i+1}({\bfla}) - k_{i}({\bfla}) )  \right) + \\
+  \frac{1}{x^{3}} \left( \frac 13 A_{i}^{(3)} + {\ve}_{1}{\ve}_{2} \left( 2 (c_{i+1}({\bfla}) - c_{i}({\bfla}) \right) - {\ve} \left( 
k_{i+1}({\bfla}) + k_{i}({\bfla}) \right) \right)  + \ldots 
\label{eq:xiexpx}
\end{multline}
where
\beq
\begin{aligned}
& A_{i}^{(k)} = \sum_{{\alpha}=1}^{N} \left( {\ac}_{i,\alpha}^{k} - \left( {\ac}_{i+1, \alpha} - {\ve} \right)^{k} \right)\\
& k_{i} ({\bfla}) = \sum_{{\alpha}=1}^{N} | {\lambda}^{(i, {\alpha})} | \\
&  c_{i} ({\bfla}) = \sum_{{\alpha}=1}^{N} \sum_{{\square} \in {\lambda}^{(i, {\alpha})}} c_{\square} \\
\label{eq:xicoeff}
\end{aligned}
\eeq

\subsection{Abelian theory}

When $N=1$ the equations \eqref{eq:tcix} read (we skip the index $\alpha$ in
the formulas for $N=1$)
\begin{multline}
\sum_{j=0}^{r} \frac{t z_{j}}{1+tz_{j}} \left( - {\ac}_{0} A_{j}^{(1)}    + {\ve}_{1}{\ve}_{2} \vev{k_{j+1} - k_{j}}  + \frac 12 \left( A_{j}^{(2)} +  (A_{j}^{(1)})^{2}  \right)\right) + \\
\sum_{0 \leq i < j \leq r} \frac{t z_{i}}{1+tz_{i}} \frac{t z_{j}}{1+tz_{j}}  (A_{i}^{(1)} - {\ve}) A_{j}^{(1)} = 0 
\end{multline}
which imply, by taking the residue at $t = - z_{i}^{-1}$:
\begin{multline}
- {\ve}_{1}{\ve}_{2} {\nabla}_{i}^{z} {\log}{\CalZ}_{1}^{\rm inst} + ( {\ac}_{i}  - {\ac}_{0} )   \left( {\ac}_{i} - {\ac}_{i+1} + {\ve} \right) + \\
\left( {\ac}_{i} - {\ac}_{i+1} \right) \sum_{j > i} \frac{z_{j}\left( {\ac}_{j} - {\ac}_{j+1} + {\ve} \right)}{z_{j}-z_{i}}  + \left( {\ac}_{i} - {\ac}_{i+1} + {\ve} \right) \sum_{i > j} \frac{z_{j}({\ac}_{j} - {\ac}_{j+1})}{z_{j}-z_{i}}   = 0 
\end{multline}
which, in turn, implies:
\beq
{\CalZ}_{1}^{\rm inst} = \prod_{0 \leq i < j \leq r} \left( 1 - z_{j}/z_{i} \right)^{\frac{\left( {\ac}_{i} - {\ac}_{i+1}  \right)\left( {\ac}_{j} - {\ac}_{j+1} + {\ve} \right)}{{\ve}_{1}{\ve}_{2}}} 
\label{eq:abelian}
\eeq

\subsection{Non-abelian theory and surface defects}

When $N=2$ the equations \eqref{eq:cixeq} (with the help of the appendix) imply the relations 
\beq
\vev{c_{i}} = \vev{{\rm polynomials\ in\ } k_{j}'s } = {\rm differential\ operator\ in}\ {\qe}_{j}'s \ {\rm on} \ {\CalZ}_{2}
\label{eq:cirel}
\eeq
The relations \eqref{eq:cirel} are not very useful, since they express unknown 
expectation values in terms of the derivatives of the partition function. 
However, when the parameters are tuned as in \eqref{eq:dege}, the measure 
\eqref{eq:cmbfla} forces the parittions ${\lambda}^{(1,{\beta})}$ to be empty when ${\beta} \neq {\alpha}$, while ${\lambda}^{(1, {\alpha})}$ is forced to have a linear structure: 
\beq
{\lambda}^{(1, {\alpha})} = ( 1^{n} ) \, , \qquad  n = | {\lambda}^{(1, {\alpha})} |
\eeq
Then: 
\beq
k_{1}({\bfla}) = n, \quad c_{1}({\bfla}) = {\ac}_{1, \alpha} n + \frac 12 {\ve}_{1}  n (n-1) 
\label{eq:cfromn}
\eeq
Physically, the linear structure of the instantons of the $U(N)_{1}$ gauge factor means that the corresponding gauge field configuration is confined ito a two-dimensional plane, where it effectively becomes a vortex. This is happening because the condition \eqref{eq:dege} makes $N$ fundamental hypermultiplets (out of $N^2$) nearly massless, thereby opening a Higgs branch. The instantons squeezed into the Abrikosov-Nielsen-Ohlesen strings  
generate a surface defect in the $A_{r-1}$ gauge theory. The gauge coupling ${\qe}_{1}$ plays the role of the two dimensional Kahler parameter of the supersymmetric sigma model living on the worldsheet of this surface defect.

\subsection{Back to the equations}
Armed with \eqref{eq:cfromn} we express:
\beq
\vev{c_{1}} = ( {\ac}_{1, \alpha} - \frac 12 {\ve}_{1} ) {\qe}_{1} \frac{d}{d{\qe}_{1}}  {\CalZ}_{2}^{\rm inst}
 + \frac 12 {\ve}_{1} \left( {\qe}_{1} \frac{d}{d{\qe}_{1}} \right)^{2} {\CalZ}_{2}^{\rm inst}
 \label{eq:cfrmnz}
 \eeq
 which makes the residue of \eqref{eq:cirel} at $t = -z_{0}^{-1}$ into a differential equation on ${\CalZ}_{2}$:
 \begin{multline}
 ( {\ve}_{1} {\nabla}_{0}^{z} )^{2} {\CalZ}_{2}^{\rm inst} \, + \,
 \left( {\mu}_{2} - {\mu}_{1} - {\ve}_{2} - \sum_{i} \frac{z_{i}A_{i}^{(1)}}{z_{i}-z_{0}}   \right) ( {\ve}_{1} {\nabla}_{0}^{z} ) {\CalZ}_{2}^{\rm inst}  \ + \\
\left[  \sum_{i} \frac{z_{i}}{z_{i}-z_{0}} \left( - {\ve}_{1}{\ve}_{2} {\nabla}_{i}^{z} + \frac 12 \left( A_{j}^{(2)} +  (A_{j}^{(1)})^{2}  \right) - {\mu}_{2} A_{i}^{(1)} \right) + \sum_{i < j} \frac{z_{i}z_{j}(A_{i}^{(1)}-{\ve})A_{j}^{(1)}}{(z_{i}-z_{0})(z_{j} - z_{0})}  \right] {\CalZ}_{2}^{\rm inst} = 0
 \label{eq:master}
 \end{multline}
 where ${\ac}_{1,\alpha} = {\mu}_{1}$, ${\ac}_{1, \beta} = {\mu}_{2}$, ${\beta}
 \in \{ 1, 2 \} \backslash \{ {\alpha} \}$. 
 Now define 
 \beq
 {\chi} (z_{0}, \ldots, z_{r} ) = z_{0}^{L_{0}}  \prod_{i=1}^{r} (z_{i}/z_{0}) ^{L_{i}} (1 - z_{i}/z_{0})^{{\delta}_{i}} \prod_{0 < i < j \leq r} (1 - z_{j}/z_{i})^{T_{ij}}\, {\CalZ}_{2}^{\rm inst}
 \label{eq:gaugerot}
 \eeq
 where
 \begin{multline}
 T_{ij} = \frac{2(m_{i} - {\ve}) m_{j} }{{\ve}_{1}{\ve}_{2}} \, , \qquad  m_{i} = \frac{A_{i}^{(1)}}{2} = {\ve} + \sum_{\alpha = 1}^{2} \frac{{\ac}_{i, \alpha} - {\ac}_{i+1, \alpha}}{2} \, , \qquad
 {\delta}_{i} = m_{i} /{\ve}_{1} \\
 L_{0} = {\Delta}_{-1} - \sum_{i=0}^{r+1} {\Delta}_{i} \, ,  \qquad
  {\ve}_{1}{\ve}_{2} {\Delta}_{i} = m_{i} ({\ve} - m_{i}), \qquad i = 1, \ldots , r \\
 {\ve}_{1}{\ve}_{2} {\Delta}_{-1} = \frac{{\ve}^{2}}{4} - \frac{({\mu}_{1} + {\ve}_{2} - {\mu}_{2})^{2}}{4} \equiv \frac{{\ve}^{2}}{4} - \frac{({\ac}_{0, 1} - {\ac}_{0, 2})^{2}}{4},  \
  {\ve}_{1}{\ve}_{2} {\Delta}_{r+1} = \frac{{\ve}^{2}}{4} - \frac{({\ac}_{r+1, 1} - {\ac}_{r+1, 2})^{2}}{4} ,\label{eq:deltamatch}
  \end{multline}
It is now straightforward to check that $\chi$ so defined solves the equations
\eqref{eq:bpz2}.  The relations \eqref{eq:deltamatch} match exactly the conjecture of \cite{AGT}. 
 
\section{The higher rank $U(n)$ theories with orbifold defects}

Let us pass to the theories with a single factor gauge group $U(n)$. We study two examples:  the $A_1$ theory or the ${\hat A}_{0}$ theory. 

\subsection{ The $A_{1}$ case.} 
The $A_{1}$ theory is the $U(n)$ gauge theory with $N_{f} = 2n$ fundamental hypermultiplets. The theory is characterized by the gauge coupling $\qe$ and $2n$ masses ${\bmt} = ({\mt}_{1}, \ldots , {\mt}_{2n})$, which are encoded in the polynomial 
\[ P(x) = \prod_{{\fe}=1}^{2n} ( x - {\mt}_{\fe} ) \]
Since the quiver consists of a single vertex, we omit the subscript
$\ib$ in ${\y}(x)$ and $P(x)$.

The fundamental $A_1$ $qq$-character is equal to 
\beq
{\x}_{1,0} (x) = {\y}(x+{\ve}) + {\qe}\ \frac{P(x)}{{\y}(x)}
\label{eq:fa1ch}
\eeq

The general $A_{1}$ $qq$-character depends on a $\tw$-tuple $\bnu$ of complex numbers, ${\bnu} = ({\nu}_{1}, \ldots , {\nu}_{\tw}) \in {\BC}^{\tw}$.  It 
is given by:
\begin{equation}
{\x}_{{\tw}, {\bnu}}(x) = 
\sum_{[{\tw}] = I \sqcup J}
{\qe}^{|J |} \prod_{i \in I, j \in J} {\bS}_{1,2}({\nu}_{i} - {\nu}_{j}) 
 \prod_{j \in J} \frac{P( x  + {\nu}_{j})}{{\y} (x + {\nu}_{j})}  \prod_{i \in  I} {\y}(x + {\ve} + {\nu}_{i}) 
\label{eq:a1nuch}
\end{equation}
It has potential poles in $\nu$'s, when ${\nu}_{i} = {\nu}_{j}$ or
${\nu}_{i} = {\nu}_{j} + {\ve}$, for $i \neq j$.

The expression \eqref{eq:a1nuch} is actually non-singular
at the diagonals ${\nu}_{i} = {\nu}_{j}$. The limit contains, however, the derivatives $\partial_{x}{\y}$. For example, for ${\wt} =2$, ${\nu}_{1}= {\nu}_{2} = 0$ the $qq$-character is equal to:
\begin{multline}
{\x}_{2,(0,0)} (x) = {\y}(x+{\ve})^{2} \left( 1 - {\qe} \frac{{\ve}_{1}{\ve}_{2}}{\ve} {\partial}_{x} \left( \frac{P(x)}{{\y}(x){\y}(x+{\ve})} \right) \right) + \\
+ 2 {\qe} P(x) \frac{{\y}(x+{\ve})}{{\y}(x)} \left( 1 - \frac{{\ve}_{1}{\ve}_{2}}{{\ve}^{2}} \right) + {\qe}^{2} \frac{P(x)^2}{{\y}(x)^2}
\label{eq:fa1ch2}
\end{multline}
The expression \eqref{eq:a1nuch}
has a first order pole at
the hypersurfaces where ${\nu}_{i} = {\nu}_{j} + {\ve}$ for some pair $i \neq j$. The residue of ${\x}_{{\tw}, {\boldsymbol{\nu}}}$ is equal to the $qq$-character ${\x}_{{\tw}-2, \boldsymbol{\nu} \backslash \{ {\nu}_{i}, {\nu}_{j} \}}$, times the polynomial in $x$ factor
\begin{equation}
\prod_{k \neq i,j} {\bS}_{1,2} ({\nu}_{k} - {\nu}_{j}) P ( x + {\nu}_{k}) \ .
\end{equation} 
The finite part ${\x}_{{\tw}, {\boldsymbol{\nu}}}^{\rm fin}$ of  the expansion of
${\x}_{{\tw}, {\boldsymbol{\nu}}}$ in $\nu_{i}$ near ${\nu}_{i} = {\nu}_{j} +{\ve}$ is the properly defined $qq$-character for the arrangement of weights $\boldsymbol{\nu}$ landing on the hypersurface ${\nu}_{i} = {\nu}_{j}+{\ve}$. It involves the terms with the derivative ${\partial}_{x}{\y}$. For example
\begin{multline}
{\x}_{2, (-{\ve},0)}^{\rm fin} = {\y}(x+{\ve}){\y}(x) + \\
+ {\qe} \left( 1 + \frac{{\ve}_{1}{\ve}_{2}}{2{\ve}^{2}} \right) P(x-{\ve})  \frac{{\y}(x+{\ve})}{{\y}(x- {\ve})} +
{\qe}P(x) \left( 1 -  \frac{{\ve}_{1}{\ve}_{2}}{\ve} \frac{{\partial}_{x}{\y}(x)}{{\y}(x)} \right) + \\
+ {\qe}^{2} \frac{P(x)P(x-{\ve})}{{\y}(x){\y}(x -{\ve})} 
\end{multline}

\subsection{ The ${\hat A}_{0}$ theory.}
The ${\hat A}_{0}$ theory (also known as the ${\CalN}=2^{*}$ theory) is characterized by one mass parameter $\mt$, the mass of the adjoint hypermultiplet, and the gauge coupling $\qe$. 

We 
give the expression for the fundamental character ${\x}_{1}(x) \equiv {\x}_{1,0}(x)$:
\begin{multline}
{\x}_{1}(x) = 
\sum_{\lambda} {\qe}^{|{\lambda}|} \ \prod_{{\square} \in {\lambda}} 
{\bS}_{1,2} ( {\mt} h_{\square} +{\ve} a_{\square} ) 
\cdot \frac{\prod_{{\square} \in {\partial}_{+}{\lambda}} {\y}(x + {\sigma}_{\square} + {\ve})}{\prod_{{\square} \in {\partial}_{-}{\lambda}} {\y}( x +
 {\sigma}_{\square})} = \\ 
= {\y} ( x + {\ve} )  \sum_{\lambda} {\qe}^{|{\lambda}|} \ \prod_{{\square} \in {\lambda}} 
{\bS}_{1,2} ( {\mt} h_{\square} +{\ve} a_{\square} ) 
\cdot \prod_{{\square} \in {\lambda} } \frac{{\y}(x + {\sigma}_{\square} - {\mt}){\y}(x + {\sigma}_{\square} + {\mt} + {\ve})}{ {\y} (x +
 {\sigma}_{\square}) {\y} (x +
 {\sigma}_{\square} + {\ve})} = \\
 = {\mathscr Y}(x+{\ve}) + {\qe} \, {\bS}_{1,2}({\mt}) \, \frac{\my{x-{\mt}} \my{x + {\ve} + {\mt}}}{\my{x}}+ \ldots
 \label{eq:cx10}
\end{multline}
Here 
\begin{equation}
{\sigma}_{\square} = {\mt}(i - j) + {\ve}(1-j)
\label{eq:cont10}
\end{equation}
is the content of ${\square}$ defined relative to the pair of weights $({\mt}, - {\mt} - {\ve})$ (cf. \eqref{eq:cont}).  
It is not too difficult to write an expression for the general ${\hat A}_{0}$ $qq$-character
${\x}_{{\wt}, {\boldsymbol{\nu}}}$, 
in terms of an infinite sum over the $\wt$-tuples of partitions,  but we feel it is not very illuminating. 

\subsection{Add surface defect}

A ${\BZ}_{p}$-type defect in the $U(n)$ gauge theory \cite{N6}
(it is the same thing as the $A_{p-1}$-type defect \cite{N4}) is specified by the choice of the coloring functions: a function $c: [n] \to {\BZ}_{p}$, assigning a $\BZ_p$ representation ${\CalR}_{c({\alpha})}$ to each color ${\alpha} = 1, \ldots , n$; a function ${\sigma}: [m] \to {\BZ}_{p}$, assigning a $\BZ_p$ representation ${\CalR}_{{\sigma}({\frak f})}$ to each flavor ${\alpha} = 1, \ldots , m = N_f$ of fundamental hypermultiplets (for the $A_1$ type theory), or an integer ${\mathfrak s} = 0, \ldots , p-1$, assigning a ${\BZ}_{p}$ representation ${\CalR}_{q}$ to the adjoint hypermultiplet (for the ${\hat A}_{0}$ theory). 

We shall also find useful to keep track of the multiplicities of colors:
\beq
{\delta}({\omega}) = \# \{ \, {\alpha} \, | \, c({\alpha})  = {\omega} \, \}
\label{eq:deltom}
\eeq

The $qq$-characters in the single node gauge theory in the presence of the ${\BZ}_{p}$-defects require, in addition to the ``evaluation parameters'' $\bnu = ({\nu}_{1}, \ldots, {\nu}_{\wt}) \in {\BC}^{\wt}$, the ${\BZ}_{p}$-coloring of the evaluation parameters ${\bfc} = ({\fc}_{1}, \ldots , {\fc}_{\wt}) \in ({\BZ}_{p})^{\wt}$. 
 
\subsubsection{The $A_1$ theory}

In the $U(n)$ theory with $2n$ fundamental hypermultiplets, we  assign a label ${\omega}$, ${\omega} = 0, 1, \ldots , p-1$ to each Coulomb parameter ${\ac}_{\alpha}$, and each fundamental mass ${\mt}_{\frak f}$:
\beq
\begin{aligned} & P(x) \ =\ \prod\limits_{{\omega}=0}^{p-1} \ P_{\omega}(x)\, , \quad P_{\omega}(x) \ =\ \prod_{{\frak f} \in [2n]\, , \ {\sigma}({\frak f}) = {\omega}} ( x - {\mt}_{\frak f} ) \\
&  {\mA}(x) \ =\ \prod\limits_{{\omega}=0}^{p-1} \ {\mA}_{\omega}(x)\, , \quad {\mA}_{\omega}(x)\ =\ \prod_{{\alpha} \in [n]\, , \ c({\alpha}) = {\omega}} ( x - {\ac}_{\alpha} ) \\
\label{eq:mushrooms}
\end{aligned}
\eeq
where
\beq
{\rm deg}\, P_{\omega}(x) \, =\,  m_{\omega}, \quad {\rm deg}\, {\mA}_{\omega} (x)\, =\, n_{\omega}\ . 
\eeq
The single $\y$-observable of the bulk theory becomes $p$ observables ${\y}_{\omega}(x)$,
\[ {\y}(x) \ =\ \prod_{{\omega}=0}^{p-1}\  {\y}_{\omega}(x) \] 
 which are $p$-periodic in $\omega$:
\[ {\y}_{{\omega} + p}(x) = {\y}_{\omega}(x) \]
The fundamental refined $qq$-characters ${\CalX}_{{\wt}, {\bnu}, {\fc}}$ of the $A_1$ theory with the ${\BZ}_{p}$-surface defect ${\CalD}_{p; c, {\sigma}}$
are given by:
\beq
{\x}_{\omega} (x) = {\y}_{{\omega}+1}(x+{\ve}) \ +\ {\qe}_{\omega}\ \frac{P_{\omega}(x)}{{\y}_{\omega}(x)}
\label{eq:qqsurfzp}
\eeq
The general refined $qq$-characters ${\x}_{{\bnu}, {\bfc}}$ of the $A_1$ theory with the ${\BZ}_{p}$-surface defect ${\CalD}_{p; c, {\sigma}}$ are given by: 
\begin{multline}
 {\x}_{{\bnu}, {\bfc}} (x) = \sum_{I \subset [{\wt}]} \ \prod_{i \in I,\, j \in [{\wt}]\backslash I} {\bS}_{+;1,2}({\nu}_{i} - {\nu}_{j})^{{\delta}_{(p)}({\fc}_{i} - {\fc}_{j})}\, {\bS}_{-;1,2}({\nu}_{i} - {\nu}_{j})^{{\delta}_{(p)}({\fc}_{i} - {\fc}_{j} + 1)} \times \\
  \times \prod_{i \in I} {\y}_{{\fc}_{i}+1}( x + {\nu}_{i} + {\ve} ) \prod_{j \in [{\wt}]\backslash I} {\qe}_{{\fc}_{j}} \frac{P_{{\fc}_{j}}(x+{\nu}_{j})}{{\y}_{{\fc}_{j}}(x+{\nu}_{j})} 
 \label{eq:a1frqch}
\end{multline}
Let us consider the special case, where ${\fc}_{i} = {\omega}$ for all $i \in [{\wt}]$. In this case \eqref{eq:a1frqch} can be written in the determinant form:
\beq
{\x}_{{\wt}, {\bnu}, {\omega}} (x) = {\Det} \, \Biggl\Vert  {\Xi}_{ij} (x) \Biggr\Vert_{i,j=1}^{\wt}
\label{eq:deta1qchn}
\eeq
with 
\beq
R_{\wt}(x) = \prod_{i=1}^{\wt} ( x - {\nu}_{i})
\eeq
and
\beq
{\Xi}_{ij}(x) = {\y}_{{\omega}+1}(x+{\nu}_{i} + {\ve}) {\delta}_{ij}\ + \ \frac{R_{\wt}({\nu}_{i}+{\ve}_{1})}{R_{\wt}^{\prime}({\nu}_{i})}\, \frac{P_{\omega}(x+{\nu}_{j})}{{\y}_{\omega}(x+{\nu}_{j})} \frac{{\qe}_{\omega}}{{\nu}_{i} - {\nu}_{j}+{\ve}_{1}}
\label{eq:xiij}
\eeq
The formula \eqref{eq:deta1qchn} is useful for computing the ${\bnu} \to 0$ limit of the refined $qq$-character. First, realize the operator 
$\Xi$  in \eqref{eq:deta1qchn} as the operator acting in the $\wt$-dimensional space $W$ of polynomials in one variable modulo those which vanish at the points ${\nu}_{1}, \ldots , {\nu}_{\wt}$:
$W = {\BC}[t]/R_{\wt}(t) {\BC}[t]$ as follows
\beq
({\Xi}(x)f)(t) = \left( {\y}_{{\omega}+1}(x + t + {\ve}) f(t) + {\qe}_{\omega} \,\frac{P_{\omega}(x+t)}{{\y}_{\omega}(x+t)} \oint_{\infty} \frac{R_{\wt}(u+{\ve}_{1}) - R_{\wt}(t)\,  }{ R_{\wt}(u) ( u + {\ve}_{1} - t )} f(u)du \right) \ {\rm mod} \ R_{\wt}(t)
\eeq
where the contour integral is taken along the large loop $|u| \to \infty$ (the subtraction of $R_{\wt}(t)$ in the numerator makes it obvious that the pole at $u+{\ve}_{1} = t$ does not contribute).  
The matrix elements \eqref{eq:xiij} are written in the basis of ``delta''-functions
(known as Lagrange interpolating polynomials)
\beq
f_{i}(t) = \frac{R_{\wt}(t)}{(t- {\nu}_{i}) R_{\wt}^{\prime}({\nu}_{i})}
\eeq
Now pass to the basis $e_{i}(t) = t^{i-1}$, $i = 1, \ldots , {\wt}$ and take a limit $R_{\wt}(t) \to t^{\wt}$ to get:  
\beq
{\x}^{[{\wt}]}_{\omega}(x) =  
{\Det} \, \Biggl\Vert \  \left( [t^{j-i}]  {\y}_{{\omega}+1}(x + t+ {\ve}) \right) \ + \ {\qe}_{\omega}\,  \sum_{k=0}^{i-1}\frac{(w-i+k)!}{k! (w-i)!} {\ve}_{1}^{k} \, \left( [t^{j-i+k}] \frac{P_{\omega}(x+t)}{{\y}_{\omega}(x+t)} \right)   \   \Biggr\Vert_{i,j=1}^{\wt}
\label{eq:deta1qchz}
\eeq

\subsubsection{The ${\hat A}_{0}$-theory, a.k.a. ${\CalN}=2^{*}$}

To write the $qq$-characters for the $U(n)$ ${\CalN}=2^*$ theory in the presence of the surface operator ${\CalD}_{p; c, q}$ we use the notations \eqref{eq:bomp}, \eqref{eq:kapp}. 

The fundamental $qq$-characters of the ${\CalN}=2^{*}$ theory with the surface operator ${\CalD}_{p; c, q}$ are given by (cf. \eqref{eq:bomp}):
\begin{multline}
 {\x}_{\omega} (x) = {\y}_{{\omega}+1}(x+{\ve}) \ 
 \sum\limits_{\lambda} \ 
{\Qfn} \, [{\lambda}]\ 
{\CalB}\, \left( \begin{matrix} {\mt} &   - {\mt} - {\ve} \\
 {\ve}_{1} & {\ve}_{2} \end{matrix} \, \biggl\vert \,  \begin{matrix}  {\mathfrak s} \\
  p \end{matrix} \, \biggr\vert\  {\bqt}\  \right)
 [{\lambda}]\ \times \\  
 \times\ \prod_{{\square} \in \lambda} \frac{{\y}_{{\omega}+ {\kappa}_{\square}  +{\mathfrak s}+1} ( x+ {\sigma}_{\square} + {\mt} + {\ve}) \, {\y}_{{\omega}+ {\kappa}_{\square}-{\mathfrak s} } ( x+{\sigma}_{\square} - {\mt})}{{\y}_{{\omega}+ {\kappa}_{\square}  +1} ( x+{\sigma}_{\square} + {\ve}) \, {\y}_{{\omega}+ {\kappa}_{\square}} ( x+ {\sigma}_{\square} )}
 \label{eq:cxomp}
 \end{multline}
 Here
\begin{equation}
{\sigma}_{\square} = {\mt}(i - j) + {\ve}(1-j)
\label{eq:cont10}
\end{equation}
is the content of ${\square}$ defined relative to the pair of weights $({\mt}, - {\mt} - {\ve})$.
 For ${\mathfrak s}=0$ the expression \eqref{eq:cxomp} simplifies to:
 \beq
 {\x}_{\omega} (x) = {\y}_{{\omega}+1}(x+{\ve}) \ 
 \sum\limits_{\lambda} \ 
{\BQ}_{\omega}^{\lambda}\, {\BB}_{\omega}^{\lambda}\  
 \prod_{{\square} \in \lambda} \frac{{\y}_{{\omega}+ {\kappa}_{\square} +1} ( x+ {\sigma}_{\square} + {\mt} + {\ve}) \, {\y}_{{\omega}+ {\kappa}_{\square}} ( x+{\sigma}_{\square} - {\mt})}{{\y}_{{\omega}+ {\kappa}_{\square}  +1} ( x+{\sigma}_{\square} + {\ve}) \, {\y}_{{\omega}+ {\kappa}_{\square}} ( x+ {\sigma}_{\square} )}
 \label{eq:cxompz}
 \eeq
 with ${\kappa}_{\square} = 1-j \, {\rm mod}(p)$. 

\subsubsection{The $qq$-characters for $U(1)$-defects in $U(n)$ theories}

The large $p$-limit of the ${\BZ}_{p}$ defect is the $U(1)$ defect, which is essentially a boundary condition in the three dimensional gauge theory with eight supercharges. 
The coloring functions $c, {\sigma}, {\mathfrak c}$ are now integer-valued.

The only change the formula \eqref{eq:a1frqch} undergoes is the replacement of the $p$-periodic Kronecker symbol by the ordinary one: ${\delta}_{(p)}(x) \to {\delta}_{x,0}$,   

\subsubsection{The $qq$-characters for surface defects in folded ${\CalN}=2^{*}$ theory}

The folded instantons is an example of the gauge origami theory \cite{N4,N5,N6}
with ${\rm rk} N_{12} = n$, ${\rm rk} N_{23} = m$, and all other
$N_{A} = 0$, $A \in \6$. 
Let us denote the Coulomb parameters of the theory by
${\ac}_{12, \alpha} = {\ac}_{\alpha}$, ${\alpha} \in [n]$, ${\ac}_{23, \beta} = {\fb}_{\beta}$, ${\beta} \in [m]$:
\beq
N_{12} = \sum_{\alpha} e^{{\ac}_{\alpha}}\, , \qquad N_{23} = \sum_{\beta} e^{{\fb}_{\beta}}
\label{eq:n123}
\eeq 
With the $\Omega$-deformation parameters ${\ve}_{a}$, $a \in \4$, the theory on the ${\BC}^{2}_{12}$ plane looks like 
the 
${\CalN}=2^{*}$ $U(n)$ theory with the mass ${\ve}_{3}$ adjoint hypermultiplet, while the theory on the ${\BC}^{2}_{23}$ plane looks like 
the 
${\CalN}=2^{*}$ $U(m)$ theory with the mass ${\ve}_{1}$ adjoint hypermultiplet. 
We have two $Y$-observables: ${\y}_{12}(x)$ and ${\y}_{23}(x)$:
\beq
{\y}_{ab}(x) \left[{\vec\lambda}\right]\  = \ \prod\limits_{\alpha} \left( \, (x - {\ac}_{ab, \alpha}) \cdot \, \prod\limits_{(i,j) \in {\lambda}^{(ab,{\alpha})}} \frac{\left( 1 + \frac{{\ve}_{b}}{x-{\ac}_{ab, \alpha} - {\ve}_{a}i - {\ve}_{b}j} \right)}{\left( 1 + \frac{{\ve}_{b}}{x-{\ac}_{ab, \alpha} - {\ve}_{a}(i-1) - {\ve}_{b}j} \right)} \right)
\label{eq:abyx}
\eeq
There are two interesting $qq$-characters we shall study, the one corresponding to the rank one theory on ${\BC}^{2}_{34}$, another corresponding to the rank one theory on ${\BC}^{2}_{14}$.

Let $N_{34} = e^{x}$. The corresponding $qq$-character
has the usual form  \eqref{eq:cx10}:
\beq
{\x}_{34}^{\rm fold}(x) 
= {\tilde\y} ( x + {\ve} )  \sum_{\lambda} {\qe}^{|{\lambda}|} \ {\mu}_{1,2}[{\lambda}]
\cdot \prod_{{\square} \in {\lambda} } \frac{{\tilde\y}(x + {\sigma}_{\square} - {\ve}_{3}){\tilde\y}(x + {\sigma}_{\square} - {\ve}_{4} )}{ {\tilde\y} (x +
 {\sigma}_{\square}) {\tilde\y} (x +
 {\sigma}_{\square} - {\ve}_{3}-{\ve}_{4})} 
 \label{eq:cxfold10}
\eeq
where ${\sigma}_{(i,j)} ={\ve}_{3}(i - 1) + {\ve}_{4}(j-1)$, and
\beq
{\mu}_{a,b}[{\lambda}] = \prod_{{\square} \in {\lambda}} 
{\bS}_{a,b} ( {\ve}_{{\tt h}(a,b)} h_{\square} +({\ve}_{a}+{\ve}_{b}) a_{\square} ) 
\eeq 
where ${\tt h}(a,b) = {\rm min}({\4}\backslash \{a, b\})$, e.g. ${\tt h}(1,2) = 3$, ${\tt h}(2,3) = 1$. 
However, the novelty is that the $\y$-function is now
\beq
{\tilde\y}(x) = {\y}_{12}(x) \prod\limits_{n=1}^{\infty} 
\frac{{\y}_{23}(x-{\ve}_{1} + {\ve}_{3}n)}{{\y}_{23}(x+{\ve}_{3}n)}
\eeq
is regularized using the infinite product representation of the $\Gamma$-function. 
Symmetrically,
\beq
{\x}_{14}^{\rm fold}(x) 
= {\hat\y} ( x + {\ve}_{2}+{\ve}_{3} )  \sum_{\lambda} {\qe}^{|{\lambda}|} \ {\mu}_{2,3}[{\lambda}]
\cdot \prod_{{\square} \in {\lambda} } \frac{{\hat\y}(x + {\tilde\sigma}_{\square} - {\ve}_{1}){\hat\y}(x + {\tilde\sigma}_{\square} - {\ve}_{4} )}{ {\hat\y} (x +
 {\tilde\sigma}_{\square}) {\hat\y} (x +
 {\tilde\sigma}_{\square} - {\ve}_{1}-{\ve}_{4})} 
 \label{eq:cxfold11}
\eeq
where ${\tilde\sigma}_{(i,j)} ={\ve}_{1}(i - 1) + {\ve}_{4}(j-1)$, and
\beq
{\hat\y}(x) = {\y}_{23}(x) \prod\limits_{n=1}^{\infty} 
\frac{{\y}_{12}(x-{\ve}_{3} + {\ve}_{1}n)}{{\y}_{12}(x + {\ve}_{1}n)}
\eeq
is regularized using the infinite product representation of the $\Gamma$-function. 
Fortunately, in the orbifold case, with the judicial choice of the coloring functions we shall not encounter infinite products. 

\subsubsection{Folding and orbifolding}

Let us now add the surface defects stretched both along the ${\BC}^{1}_{1}$ and the ${\BC}^{1}_{3}$ lines, by performing the
orbifold $(z_2 , z_4) \mapsto ({\varpi}_{p}z_{2}, {\varpi}_{p}^{-1}z_{4})$. The $qq$-characters \eqref{eq:cxfold10}, \eqref{eq:cxfold11} fractionalize:
\beq
\begin{aligned}
& {\x}_{34, \omega}^{\rm fold}(x) 
= 
{\tilde\y}_{\omega +1} ( x + {\ve}_{1}+{\ve}_{2} )  \sum_{\lambda} Q_{\lambda}^{{\omega}, {\BZ}_{p}} \ {\mu}_{1,2}^{{\BZ}_{p}}[{\lambda}]
\cdot \prod_{{\square} \in {\lambda} } \frac{{\tilde\y}_{{\om}+{\kappa}_{\square}}(x + {\sigma}_{\square} - {\ve}_{3}){\tilde\y}_{{\om}+{\kappa}_{\square}+1}(x + {\sigma}_{\square} - {\ve}_{4} )}{ {\tilde\y}_{{\om}+{\kappa}_{\square}} (x +
 {\sigma}_{\square}) {\tilde\y}_{{\om}+{\kappa}_{\square}+1} (x +
 {\sigma}_{\square} - {\ve}_{3}-{\ve}_{4})}  \\
 & {\x}_{14, \omega}^{\rm fold}(x) 
= 
{\hat\y}_{\omega +1} ( x + {\ve}_{2}+{\ve}_{3} )  \sum_{\lambda} Q_{\lambda}^{{\omega}, {\BZ}_{p}} \ {\mu}_{3,2}^{{\BZ}_{p}}[{\lambda}]
\cdot \prod_{{\square} \in {\lambda} } \frac{{\hat\y}_{{\om}+{\kappa}_{\square}}(x + {\tilde\sigma}_{\square} - {\ve}_{1}){\hat\y}_{{\om}+{\kappa}_{\square}+1}(x + {\tilde\sigma}_{\square} - {\ve}_{4} )}{ {\hat\y}_{{\om}+{\kappa}_{\square}} (x +
 {\tilde\sigma}_{\square}) {\hat\y}_{{\om}+{\kappa}_{\square}+1} (x +
 {\tilde\sigma}_{\square} - {\ve}_{1}-{\ve}_{4})}  \\
 \end{aligned}
 \label{eq:cxffr10}
\eeq
where
\beq
Q_{\lambda}^{{\omega}, {\BZ}_{p}} = \prod_{(i,j) \in {\lambda}} {\qe}_{\omega +1 - j}\, , \ 
\eeq
and, for $(a,b) = (1,2)$ or $(2,3)$
\beq
{\mu}_{a,b}^{{\BZ}_{p}}[{\lambda}]
 = \prod_{{\square} \in {\lambda}} 
{\bS}_{+, a,b} \left( {\ve}_{{\tt h}(a,b)} h_{\square} +({\ve}_{a}+{\ve}_{b}) a_{\square} \right)^{{\delta}_{p}(a_{\square})} 
{\bS}_{-, a,b} \left( {\ve}_{{\tt h}(a,b)} h_{\square} +({\ve}_{a}+{\ve}_{b}) a_{\square} \right)^{{\delta}_{p}(a_{\square}+1)} 
\label{eq:muabp}
\eeq

\subsubsection{More fun with folded theories}

Let us present here the $A_1$ analogue of the folded theory. 
By that we mean the theory which looks like the $A_1$ type theory on the ${\BC}^{2}_{12}$ plane.  We get it by performing
the ${\BZ}_{3}$ orbifold of the folded setup $n {\BC}^{2}_{12} + m {\BC}^{2}_{23}$ we discussed so far, where the group ${\BZ}_{3}$ acts in the ${\BC}^{2}_{34}$ plane: $(z_{3}, z_{4}) \mapsto ({\varpi}z_{3}, {\varpi}^{-1}z_{4})$. The theory now has three fractional couplings ${\qe}_{0}, {\qe}_{1}, {\qe}_{2}$. 
We send ${\qe}_{1}, {\qe}_{2} \to 0$. 

There are six Chan-Paton spaces now, 
$N_{A, \omega}$, $A = 12, 23$, $\omega = 0,1,2$. From the point of view of the ${\BC}^{2}_{12}$ observer, $N_{12,0}$
is the color space, while $N_{12, 1}$, $N_{12,2}$ are 
the multiplicities of the fundamental hypermultiplets. 
This is the content of the $A_1$ theory. 

Life is more intricate on the ${\BC}^{2}_{23}$ plane. Because of the ${\qe}_{1}, {\qe}_{2} \to 0$ limit the instantons there can only ``grow'' in the ${\BC}^{1}_{2}$ direction, i.e. this theory is effectively two dimensional. Moreover, 
\beq
S_{23, 0} = N_{23,0} - P_{2} K_{23,0}\, , \ S_{23, 1} = N_{23, 1} + q_{3} P_{2} K_{23, 0}\, , \ S_{23, 2} = N_{23, 2} 
\eeq
with 
\beq
P_{2} K_{23, 0} = \sum_{\beta} e^{b_{\beta}} ( 1- q_{2}^{d_{\beta}} ) 
\eeq 
where $N_{23,0} = \sum_{\beta} e^{b_{\beta}}$ and $d_{\beta}$ are the vortex fluxes. 

We can therefore view the combined theory as a surface defect in the $A_1$ type theory. In particular, by choosing
$N_{23,0} = e^x$, $N_{23,1} = N_{23,2} = 0$ to be rank one, the resulting operator is identified with the ${\ve}_{1}$-deformed
${\tilde\bQ}(x)$, the ``second solution'' of the $T-Q$-equation
\beq
\left\langle \frac{{\bQ}(x+{\ve}_{1} + {\ve}_{2})}{{\bQ}(x+{\ve}_{1})} + {\qe} P(x) \frac{{\bQ}(x-{\ve}_{2})}{{\bQ}(x)} \right\rangle = T(x)
\eeq
If, on the other hand, we choose $N_{23,0} = 0 = N_{23,2}$, and $N_{23,1} = e^{x}$, the resulting surface defect will be
the Baxter operator ${\bQ}(x)$ itself. See \cite{NP2, N3,N4, N5} for more details on Baxter operators in gauge theory. 

\subsection{KZ-type equations}

In this section we demonstrate that the differential equations from the world of two dimensional conformal field theories, such as the KZ equations \cite{Knizhnik:1984} and KZB \cite{Bernard:1988} equations are verified by the regular surface defect partition functions of the above supersymmetric gauge theories.

In fact, we shall get the special case of the KZ and KZB equations. For the $A_1$ theory we'll get the KZ equation for the four-point conformal block, with two generic vertex operators and two minimal ones. For the ${\hat A}_{0}$ case we'll  get the KZB equation which is obeyed by the one-point conformal block on the torus of a minimal vertex operator in $SU(n)$ WZW theory, corresponding to the minimal orbit of $SU(n)$. Such equations (in genus one) were studied in \cite{Etingof:1993gk}. Of course, the unitary WZW theory vertex operators have discrete labels. The level $k$ of the WZW model is also quantized. Our equations have continuous parameters, thus extending the range of natural parameters of the two dimensional CFT into the complex domain.

 Throughout this section we use the notation ${\ve}_{3} = {\mt}$. 
 
 We shall assume the $c$-coloring function of our surface defect to be given by:
\beq
c ({\alpha} ) = {\alpha}, \qquad {\alpha} = 1, \ldots ,  n
\label{eq:ccol}
\eeq
The ${\y}_{\omega}(x)$-observable in the presence of such surface operator on the instanton configuration $\bla$ evaluates to 
\beq
{\y}_{\omega}(x)[{\bla}] = ( x - {\ac}_{\omega} ) \prod_{{\beta}=1}^{n} \prod_{{\square}\in {\lambda}^{({\beta})}} \left( \frac{x - {\ac}_{\beta} - c_{\square} - {\ve}_{1}}{x - {\ac}_{\beta} - c_{\square}}  \right)^{{\delta}_{n}( {\beta} + j-1 - {\omega})} \left( \frac{x - {\ac}_{\beta} - c_{\square} - {\ve}_{2}}{x - {\ac}_{\beta} - c_{\square}- {\ve}}  \right)^{{\delta}_{n}( {\beta} + j - {\omega})}
\eeq
The strategy is to look at the coefficient $[ x^{-1} ] {\x}_{\omega}(x)$ in the orbifold $qq$-characters ${\x}_{\omega}(x)$, ${\omega} = 0, \ldots, n-1$:
\beq
[ x^{-1} ] \, \vev{{\x}_{\omega}(x)} = 0 
\label{eq:maineq}
\eeq 
To compute \eqref{eq:maineq}, we need to know the large $x$ expansion of ${\y}_{\omega}(x)$:
\begin{multline}
{\y}_{\omega}(x) = ( x - {\ac}_{\omega} )\, {\exp} \ \left( \frac{{\ve}_{1}}{x}  {\nu}_{{\om}-1}  + \frac{{\ve}_{1}{\ve}_{2}}{x^{2}} k_{{\omega}-1} + \frac{{\ve}_{1}}{x^{2}} \left( {\sigma}_{{\omega}-1} - {\sigma}_{\omega} \right)   + \ldots \right) = \\
x - {\ac}_{\omega} + {\ve}_{1} {\nu}_{{\om}-1} 
+ \frac{{\ve}_{1}}{x} \, \left( {\sigma}_{{\omega}-1} - {\sigma}_{\omega} - {\ac}_{\omega}  {\nu}_{{\omega}-1}  + {\ve}_{2} k_{\omega -1} + \frac{{\ve}_{1}}{2} {\nu}_{{\omega}-1}^2 \right)  
\label{eq:yomexp}
\end{multline}
where
\begin{multline}
k_{\omega} = \# {\CalK}_{\omega} \ , \qquad  {\nu}_{\om} = k_{\om} - k_{{\om}+1} \ , \qquad {\sigma}_{\omega} = \frac{{\ve}_{1}}{2} k_{\omega} + \sum_{({\beta}, {\square}) \in {\CalK}_{\omega}} \left( {\ac}_{\beta} + c_{\square} \right) \\
{\CalK}_{\omega}  = \left\{ ( {\beta}, {\square} ) \, | \, {\beta} \in [n], \ {\square} = (i,j) \in {\lambda}^{({\beta})}, \ {\beta} + j-1 \equiv {\omega}\, {\rm mod}(p) \right\}  \, ,  
\end{multline}

\subsubsection{The $A_1$ case} We take the matter coloring function to be
\beq
{\sigma}({\mathfrak f})  = [({\mathfrak f}-1)/2]+1, \qquad {\mathfrak f} = 1, \ldots , 2n
\label{eq:mcol}
\eeq
The $qq$-characters are
\beq
{\x}_{\omega}(x) = {\y}_{{\omega}+1} ( x + {\ve}) + {\qe}_{\omega} P_{\omega}(x) {\y}_{\omega}^{-1}(x) , \qquad {\omega}  = 0, \ldots , n-1
\label{eq:qqchp}
\eeq
with
\beq
P_{\omega}(x) = (x + {\mt}_{2{\omega}})(x + {\mt}_{2{\omega}+1})
\eeq
Using \eqref{eq:yomexp} we derive:
\begin{multline}
\frac{1}{\ve_1} [ x^{-1} ] {\x}_{\omega}(x) = \, {\CalD}_{\omega} - {\qe}_{\omega} {\CalD}_{{\omega}-1} + 
\frac{{\ve}_{1}}{2} \left(   {\nu}_{\om}^2  
+ {\qe}_{\omega} {\nu}_{{\omega}-1}^2    \right) - \\
- {\qe}_{\omega} \left( {\ac}_{\omega} + {\mt}_{2{\omega}} + {\mt}_{2{\omega}+1} \right) {\nu}_{{\omega}-1} - {\ac}_{\omega +1 }  {\nu}_{\om} + {\qe}_{\omega} \frac{P_{\omega}({\ac}_{\omega})}{{\ve}_{1}} \\
\end{multline}
where
\beq
{\CalD}_{\omega} =  {\sigma}_{{\omega}} - {\sigma}_{\omega + 1}   + {\ve}_{2} k_{\omega} \eeq
Thus, we derived a system of equations, relating $\vev{{\CalD}_{\omega}}$ to the derivatives of the surface defect partition function ${\bf\Psi}$:
\begin{multline}
\vev{{\CalD}_{\omega} - {\qe}_{\omega} {\CalD}_{{\omega}-1}} = \\
- \frac 12 {\ve}_{1} \vev{{\nu}_{\om}^2}  
- \frac 12 {\ve}_{1} {\qe}_{\omega} \vev{{\nu}_{{\om}-1}^2} +  
{\qe}_{\omega} \left( {\ac}_{\omega} + {\mt}_{2{\omega}} + {\mt}_{2{\omega}+1} \right) \vev{{\nu}_{{\omega}-1}} + {\ac}_{\omega +1 } \vev{{\nu}_{\om}} -  {\qe}_{\omega} \frac{P_{\omega}({\ac}_{\omega})}{{\ve}_{1}} \vev{1}\, ,  \\
{\omega} = 0, \ldots, n-1
\end{multline}
which becomes a second order differential equation on ${\bf\Psi}$ once we use the obvious relation
\beq
\sum_{{\om}=0}^{n-1} \vev{{\CalD}_{\omega}} = {\ve}_{2} {\mathscr{D}}^{\qe}
{\bf\Psi} 
\eeq
In \cite{NT} we map this equation to the KZ equation for the special four-point genus zero conformal blocks of the $SL(n)$ current algebra. 

\subsubsection{The regular defects in the ${\CalN}=2^*$ theory} 

We take ${\mathfrak s}=0$. Let us first consider the case of the regular surface defect, i.e. the one for which ${\delta}({\omega}) = 1$ for all $\omega \in {\BZ}_{p}$. 

The large $x$-expansion of  the fundamental ${\BZ}_{n}$-refined $qq$-character 
${\x}_{\omega}(x)$ given by \eqref{eq:cxompz}
 is equal to:
 \begin{multline}
 \frac{1}{{\BB}_{\omega} }\, {\x}_{\om}(x) =  x + {\ve} - {\ac}_{c^{-1}({\om} +1)} + {\ve}_{1}  {\nu}_{\omega} + \\
 \frac{1}{x}  \, \left(  \frac 12 \left( {\ve}_{1} {\nu}_{\om} - {\ac}_{c^{-1}({\om} +1)} \right)^{2} - \frac 12 {\ac}_{c^{-1}({\om} +1)}^2  + 
 {\ve}_{1}   {\CalD}_{\omega}    - {\mt} \sum_{{\omega}'=0}^{p-1} \left( \left( {\mt} + {\ve} \right) {\nabla}_{\omega'}^{\qe} + \left( {\ve}_{1} {\nu}_{\omega'} - {\ac}_{c^{-1}({\om}'+1)} \right) {\nabla}^{z}_{\om'} \right) {\log}{\BB}_{\omega}    \right)  + \ldots
 \label{eq:rfcex}
 \end{multline}
 It implies the relation between the expectation value of ${\CalD}_{\omega}$ and
 the differential operator acting on the partition function $\tilde{\bf\Psi} \equiv {\bf\Psi}^{\rm inst}_{{\hat A}_{0}, {\BZ}_{p}} ({\ba}; {\mt}; {\ept}; {\bqt})$ of the regular surface defect. 
 \begin{multline}
0 = \frac{1}{{\BB}_{\omega}}\, [x^{-1}]\vev{{\x}_{\omega} (x)} = {\ve}_{1} \vev{{\CalD}_{\omega}} {\tilde{\bf\Psi}} + \frac 12 ({\ve}_{1}{\nabla}_{\omega}^z)^{2} {\tilde{\bf\Psi}} - {\ve}_{1}{\ac}_{c^{-1}({\om}+1)} {\nabla}_{\omega}^{z} {\tilde{\bf\Psi}}  - \\
 {\mt} \sum_{\omega' = 0}^{p-1} \left( \left( {\mt} + {\ve} \right) {\tilde{\bf\Psi}}\, {\nabla}_{\omega'}^{\qe} + \left( {\ve}_{1} {\nabla}^{z}_{\omega'} {\tilde{\bf\Psi}} - {\ac}_{c^{-1}({\om}'+1)} {\tilde{\bf\Psi}} \right) {\nabla}^{z}_{\omega'} \right) \,
 {\log}( {\BB}_{\omega}  )  
 \label{eq:mcxom}
 \end{multline} 
Sum over $\om$ to get the linear differential equation (using some identities from the appendix)
\beq
 \left( p{\ve}_{1}{\ve}_{2} {\Nq} +  {\ve}_{1} \left( {\ve}_{2} {\rv} - {\ba}^{+}_{s} \right) \cdot  {\nabla}^{\bzv}   + \frac 12 {\ve}_{1}^{2} {\Delta}_{\bzv}  + \frac 12 {\mt}({\mt}+{\ve}_{1}) {\Delta}_{\bzv} {\log} ({\tilde\BDe}) + ({\mt}+{\ve}_{1})^2 {\hat u} \right) {\hat{\bf\Psi}}  = 0 
 \label{eq:kzb1}
\eeq
on the normalized partition function
\beq
{\hat{\bf\Psi}} =  {\tilde\BDe}^{1 + {\mt}/{\ve}_{1}} \cdot {\tilde{\bf\Psi}}
\eeq
where ${\ba}^{+}_{s}  \cdot  {\nabla}^{\bzv} = \sum_{\om} {\ac}_{c^{-1}({\om}+1)} {\nabla}_{\om}^{z}$ and  (cf. \eqref{eq:dbbbt})
\beq
{\hat u} = {\BDe}^{-1} \left(  p {\Nq} + {\rv} \cdot {\nabla}^{\bzv} - \frac 12 {\Delta}_{\bzv} \right) {\BDe} - p\frac{\ve_2}{{\mt}+{\ve}_1} {\Nq}{\chi}
\label{eq:upotreg}
\eeq
actually vanishes, ${\hat u} \equiv 0$, as we now demonstrate using  the sub-regular defects.

\subsection{The sub-regular defects in the ${\CalN}=2^{*}$ theory} 

The sub-regular defect is the one where the multiplicities ${\delta}({\omega})$ of 
the ${\BZ}_{p}$-representations occuring in the coloring of the Chan-Paton space are $0$ and $1$ with at least one $0$. 

The fundamental refined $qq$-characters ${\x}_{\omega}(x)$ behave as $x^{{\delta}({\omega}+1)}$ when $x \to \infty$. We shall have to expand $\x_{\omega}$ with ${\delta}({\omega} +1) = 0$ up to the $x^{-2}$ terms. The ${\x}_{\omega}$ with 
${\delta}( {\omega} +1) = 1$ are expanded to the $x^{-1}$ order, as in the regular case. 

{}Let us denote by ${\tilde{\bf\Psi}}_{\delta}$ the instanton part of the partition function of the surface defect characterized by the function $\delta$.
Note that when ${\delta} = {\delta}_{\omega_0}$ is supported at only one value of $\omega$, i.e. ${\delta}_{\omega_0} ({\omega}) = {\delta}_{\omega, \omega_0}$, this partition function coincides with
our friend \eqref{eq:bombtom}
\beq
{\tilde{\bf\Psi}}_{\delta_{\omega_0}} = {\tilde\BB}_{\omega_0} \equiv
{\tilde\BB}_{p,\omega_{0}} ({\bar\ve}; {\bqt})
\label{eq:normsubreg}
\eeq
 Let us denote by ${\nabla}_{\delta}^{z}, {\nabla}_{\delta}^{\qe}$ the operators, cf. \eqref{eq:bbbt} 
\beq
{\nabla}_{\delta}^{z} = \sum_{{\omega}=0}^{p-1} {\delta}({\omega} +1) {\nabla}_{\omega}^{z}, \qquad {\nabla}_{\delta}^{\qe} = \sum_{{\omega}=0}^{p-1} {\delta}({\omega} +1) {\nabla}_{\omega}^{\qe}\, , 
\label{eq:delnab}
\eeq
The equations $[x^{-1}] \vev{{\x}_{\omega}(x)} = 0$ give:
\beq
{\ve}_{1} {\nabla}_{\omega}^{z} {\log}{\tilde{\bf\Psi}}_{\delta} + 
{\mt} {\nabla}_{\delta}^{z} {\log} {\BB}_{\omega}  = 0 , \qquad {\rm when} \ {\delta}({\omega +1}) = 0
\label{eq:1subreg}
\eeq
which imply \eqref{eq:diffrel1}, with the help of \eqref{eq:normsubreg}. Now, using \eqref{eq:bfrobom}, we get:
\beq
{\nabla}_{\om}^{z} {\hat\Psi}_{\delta}  = 0 , \qquad {\delta}({\om}+1) = 0, 
\eeq
where
\beq
{\hat\Psi}_{\delta} = {\tilde\BB}_{\delta}^{-1} {\tilde{\bf\Psi}}_{\delta} , \qquad {\tilde\BB}_{\delta}
 = \prod_{\om} {\tilde\BB}_{\om}^{{\delta}_{\om}}
\label{eq:normpsid}
\eeq
In turn, the equations $\frac{1}{{\BB}_{\om}} [x^{{\delta}({\omega} +1) -2} ] \vev{{\x}_{\omega}(x)} = 0$   produce upon summing over 
$\omega \in {\BZ}_{p}$ while using \eqref{eq:1subreg}:
\beq
p\, {\ve}_{1}{\ve}_{2} {\qe} \frac{\partial}{\partial\qe} {\hat\Psi}_{\delta} + 
 \left[   \sum_{\omega} \frac 12 \left({\ve}_{1} {\nabla}_{\omega}^{z} \right)^{2}  +  \left( {\ve}_{2} {\rho}_{\om} - {\ac}_{c^{-1}({\om} +1)} \right) {\ve}_{1} {\nabla}_{\om}^{z} \right] {\hat\Psi}_{\delta} + {\mathfrak {\tilde U}}_{\delta} ({\bzv}; {\qe})   {\hat\Psi}_{\delta} = 0
 \label{eq:2subreg}
 \eeq
 with
 \beq
 \begin{aligned}
 {\mathfrak {\tilde U}}_{\delta} ({\bzv}; {\qe})  \ =\ &
 \left( {\ve}_{1}{\ve}_{2} \left( p {\Nq} + {\rv}\cdot \nabla^{\bzv}  \right) + \frac 12 {\ve}_{1}^{2} {\Delta}_{\bzv} \right) \, {\log}( {\tilde\BB}_{\delta} )
 \\
  + &  {\ve}_{1}^{2} \sum_{\om} {\nabla}^{z}_{\om} {\log}({\tilde\BB}) \left( \frac 12 {\delta}_{\om +1} {\nabla}_{\om}^{z} {\log}({\tilde\BB}) - 
  {\nabla}^{z}_{\om} {\log}({\tilde\BB}_{\delta}) \right) \\
  + & \frac 12 {\ve}_{1} {\nabla}^{z}_{\delta} \left( {\mt} {\nabla}_{\delta}^{z}  + {\mt}+ 2 {\ve} \right) {\log}({\tilde\BB})  \\
   & \qquad - {\mt}({\mt} + {\ve}) {\nabla}_{\delta}^{\qe} {\log}({\BB}) - {\mt} \sum_{\omega} \vev{{\tilde{\CalD}^{-}}_{\delta}}_{\om} 
 \label{eq:3subreg}
 \end{aligned}
 \eeq
Here $\vev{{\tilde\CalD}_{\delta}^{-}}_{\omega}^{~}$  is the average of the observable
 \beq
 {\tilde\CalD}_{\delta}^{-} = \sum_{\omega'} {\delta}({\omega}' +1) \left( {\tilde c}_{\omega'} - {\tilde c}_{\omega'+1}  - \frac{\mt}{2} \left( {\tilde k}_{\omega'} - {\tilde k}_{\omega' + 1} \right) \right) 
 \label{eq:cddm}
 \eeq
 with respect to the measure:
 \beq
 \vev{{\tilde\CalD}_{\delta}^{-}}_{\omega}^{~} = \frac{1}{{\BB}_{\omega}} \sum_{\lambda}\,  {\tilde\CalD}_{\delta}^{-} \, {\BQ}_{\omega}^{\lambda}\, {\BB}_{\omega}^{\lambda}
 \label{eq:vvcddm}
 \eeq
 where 
 \[
 {\tilde k}_{\omega'}[{\lambda}] = \# \{ \, {\square}  = (i,j) \, | \, {\square} \in {\lambda}, \ {\omega} + 1  - j \equiv \, {\omega}' \ {\rm mod}\ p \, \} \ , \]
 so that
 \[ {\BQ}_{\omega}^{\lambda} = \prod_{\omega'}\  {\qe}_{\omega'}^{\ {\tilde k}_{\omega'}[{\lambda}]} \ , \]
 and
 \[ 
 {\tilde c}_{\omega'}[{\lambda}] = \sum_{{\square} \in {\lambda}, \ {\omega} + 1  - j \equiv \, {\omega}' \ {\rm mod}\ p} \, {\sigma}_{\square} \]
It looks like we are not gaining much as the expression \eqref{eq:3subreg} looks even worse then \eqref{eq:kzb1}, \eqref{eq:upotreg}, since it contains an unknown quantity 
\beq
 \sum_{\omega} \vev{{\CalD}_{\delta}^{-}}^{~}_{\omega} \ . \label{eq:ddel}
 \eeq  
However, the summit is close. We know that ${\hat{\bf\Psi}}_{\delta_{\om}} = 1$ for any ${\om} \in {\BZ}_{p}$. Therefore, 
\beq {\mathfrak {\tilde U}}_{\delta_{{\om}}} ({\bzv}; {\qe}) = 0 \, , \quad
{\om} \in {\BZ}_{p} \label{eq:upotom0}
\eeq
We can use this to simplify \eqref{eq:3subreg} quite considerably: 
\[ {\mathfrak {\tilde U}}_{\delta}({\bzv}; {\qe}) = {\mathfrak {\tilde U}}_{\delta}({\bzv}; {\qe}) - \sum_{\om \in {\BZ}_{p}}\ {\delta} ({\omega}+1)\, {\mathfrak {\tilde U}}_{{\delta}_{\om}}({\bzv}; {\qe}) \] 
which removes all the terms linear in $\delta$, including \eqref{eq:ddel}, leading to
\beq
{\mathfrak {\tilde U}}_{\delta}({\bzv}; {\qe}) = - \frac 12 {\mt} ( {\mt} + {\ve}_{1}) \sum_{{\om}' \neq {\om}''} {\delta}_{{\om}'+1} {\delta}_{{\om}''+1} \, {\nabla}_{\om'}^{z} {\nabla}_{\om''}^{z} {\log} ({\tilde\BDe}) 
\label{eq:33subreg}
\eeq
Now sum \eqref{eq:upotom0} over ${\om} \in {\BZ}_{p}$ using the expression \eqref{eq:3subreg}. The terms containing ${\tilde\CalD}_{\delta_{\om_0}}$ cancel out, leaving us with the identity:
 \begin{multline}
0  =
 \left( p {\Nq} + {\rv}\cdot \nabla^{\bzv}  \right)  \left( {\ve}_{1}{\ve}_{2} {\log}({\tilde\BB}) - {\mt} ({\mt}+{\ve}) {\log} ({\BB}) \right) +  \\
 + \frac 12 {\ve}_{1} ({\mt} + {\ve}_{1} ) {\Delta}_{\bzv} {\log}{\tilde\BB}  
 - \frac 12  {\ve}_{1}^{2} \sum_{\om}  \left( {\nabla}^{z}_{\om} {\log}({\tilde\BB}) \right)^{2}  
 \label{eq:4subreg}
 \end{multline}
which, upon the substitutions  \eqref{eq:bbbtddt}, \eqref{eq:dbbbt} is equivalent to ${\hat u}= 0$, where $\hat u$ is given by \eqref{eq:upotreg}.
Use the trigonometric limit \eqref{eq:bomtrig}, \eqref{eq:btomtrig} as the initial condition in the heat equation to prove the Eq. \eqref{eq:detdeth}. 
Now, armed with \eqref{eq:detdeth}, even without exact knowledge of what the function $\chi ({\qe})$ is, we conclude:
\beq
{\mathfrak {\tilde U}}_{\delta}({\bzv}; {\qe}) = - \frac 12 {\mt} ( {\mt} + {\ve}_{1}) \sum_{{\om}' \neq {\om}''} {\delta}_{{\om}'+1} {\delta}_{{\om}''+1} \, {\nabla}_{\om'}^{z} {\nabla}_{\om''}^{z} {\log} ({\theta}_{11}(z_{\om'}/z_{\om''} ; {\tau})) 
\label{eq:33subreg}
\eeq
This concludes the derivation in the ${\CalN}=2^{*}$ case.
The equation obeyed by the regular defect ${\delta}({\omega}) = 1$, can be now written quite explicitly (we restored the
notation ${\mt} = {\ve}_{3}$):
\beq
 \left( n{\ve}_{1}{\ve}_{2} {\Nq} + \frac 12 {\ve}_{1}^{2} {\Delta}_{\bzv}  - {\ve}_{3} ( {\ve}_{3} + {\ve}_{1}) \sum_{{\alpha} < {\beta}}  \, {\wp} \left( w_{\alpha}/w_{\beta} ; {\tau} \right) \right) {\Psi}  = 0 
 \label{eq:kzb2}
\eeq
where
\beq
{\Psi} = \prod\limits_{\alpha} \ w_{\alpha}^{\frac{p_{\alpha}}{{\ve}_{1}}} \, {\hat\bf\Psi}
\eeq
and
\beq
w_{\alpha} = z_{c({\alpha})-1}\, , \qquad p_{\alpha} = {\ac}_{\alpha} - {\ve}_{2} {\rho}_{c({\alpha})-1}
\eeq
The equation \eqref{eq:kzb2} is the KZB equation for the special one-point conformal block \cite{Etingof:1993gk}. This is one of the strongest confirmations of the original conjecture \cite{N1}.

A couple of remarks are in order. 

$\bullet$ The theory in the bulk is actually invariant under the permutation ${\ve}_{3} \leftrightarrow {\ve}_{4}$. However the surface defect is not, since with our choice ${\mathfrak s}=0$ 
the transversal $z_4$ direction is twisted by the orbifold group ${\BZ}_{p}$
while the $z_{3}$ is neutral. 
$\bullet$
In the limit to the pure ${\CalN}=2$ theory, ${\mt} \to \infty$, ${\qe} \to 0$ with ${\mt}^{2n} {\qe} = {\Lambda}^{2n}$ finite we recover the result of \cite{Braverman:2004vv}\footnote{With the additional bonus of providing not only an equation, but also its solution}, while in the purely perturbative limit ${\qe} \to 0$ with ${\mt}$ finite we'll get the results of \cite{Braverman:2010ei}. 
Also, our result proves for all $n$ the conjecture made for $n=2$ in \cite{AT}. 

$\bullet$ 
It is not much work to extend this calculation in the folded case. 
We take $p = n+m$ and choose the coloring function $c$ so that
the $c(12, {\alpha}) \neq c(23, {\beta})$ for ${\alpha} \in [n]$, ${\beta} \in [m]$. 
The regular surface defect is solving the generalization of
\eqref{eq:kzb2}:
\begin{multline}
  (m+n) {\ve}_{2} {\Nq} {\Psi} + \frac 12 {\ve}_{1} \sum_{\alpha \in [n]} {\nabla}_{w_{\alpha}}^{2} {\Psi} + \frac 12 {\ve}_{3} \sum_{\beta \in [m]} {\nabla}_{u_{\beta}}^{2}  {\Psi}  - \\
- ( {\ve}_{3} + {\ve}_{1}) \cdot  \left(  \frac{{\ve}_{3} }{{\ve}_{1}} \sum_{{\alpha}' < {\alpha}''}  \, {\wp} \left( w_{\alpha'}/w_{\alpha''} ; {\tau} \right)   + \frac{{\ve}_{1}}{{\ve}_{3}} \sum_{{\beta}' < {\beta}''}  \, {\wp} \left( u_{\beta'}/u_{\beta''} ; {\tau} \right) +  \sum_{{\alpha},  {\beta}}  \, {\wp} \left( w_{\alpha}/u_{\beta} ; {\tau} \right)
 \right) {\Psi}  = 0 
 \label{eq:kzb3}
\end{multline}
where $w_{\alpha} = z_{c(12, {\alpha})-1}$, $u_{\beta} = z_{c(23,{\beta}-1)}$. 
 
It would be nice to compare these equations to the KZB
equations for the one-point genus one conformal block for the super-Kac-Moody algebra $\hat{\frak{sl}(n|m)}$. 
It appears that we have found the elliptic generalization of the super-Calogero system \cite{sergeev}.

\section{Bethe/gauge correspondence}

In the limit ${\ve}_2 \to 0$ our theories become effectively
${\CalN}=(2,2)$ two dimensional with ${\BC}^{1}_{2}$ as a worldsheet. Such theories are in correspondence with the
quantum integrable systems \cite{NS1, NS2, NS3}. 
The ${\BZ}_{p}$ orbifold creates a point-like singularity in the ${\BC}^{1}_{2}$ space, which is the codimension two defect in that effective two dimensional theory. It can be represented by the operator ${\bf\Psi}(z)$ in the twisted chiral ring which depends on the additional fractional couplings $z_{\omega}$. As is well-known, the ${\ve}_{2} \to 0$ limit of \eqref{eq:kzb2} becomes the eigenvalue problem:
\beq
{\Psi} (w, {\ac}, {\mt}, {\tau}; {\ve}_{1}, {\ve}_{2}) = e^{\frac{1}{{\ve}_{2}} \, {\tilde \CalW} ({\ac}, {\mt}, {\tau}; {\ve}_{1})} \cdot \left( {\chi} (w, {\ac}, {\mt}, {\tau}; {\ve}_{1}) + \ldots \right) 
\eeq
where $\ldots$ denote the terms which vanish in the $\ve_2 \to 0$ limit\footnote{it is important that $\ac$ etc. are generic here, see
\cite{Jeong:2017pai} for the first steps in the study of non-generic cases}, and (we return to the ${\mt} = {\ve}_{3}$ notation):
\beq
\left( - \frac 12 {\ve}_{1}^{2} {\Delta}_{\bzv}  + {\mt} ( {\mt} + {\ve}_{1}) \sum_{{\alpha} < {\beta}}  \, {\wp} \left( w_{\alpha}/w_{\beta} ; {\tau} \right) \right) {\chi} = E ({\ac}, {\mt}, {\tau}; {\ve}_{1}) {\chi} 
\label{eq:eigv}
\eeq
where
\beq
E = {\qe} \frac{\partial}{\partial\qe} {\tilde \CalW} ({\ac}, {\mt}, {\tau}; {\ve}_{1})
\eeq
and
\beq
{\tilde \CalW} ({\ac}, {\mt}, {\tau}; {\ve}_{1}) = {\rm lim}_{{\ve}_{2} \to 0} \, {\ve}_{2} {\rm log}{\CalZ}({\ac}, {\mt}, {\tau}; {\ve}_{1},{\ve}_{2})
\label{eq:twsup}
\eeq
is the effective twisted superpotential \cite{NS3}.

The limit ${\chi} (w, {\ac}, {\mt}, {\tau}; {\ve}_{1})$ is the so-called Jost function of the elliptic Calogero-Moser system.

$\bullet$

In the non-generic case ${\mt}/{\ve}_{1} \in {\BZ}$ the paper
\cite{Felder:1995iv} gives an integral representation
for the solution of \eqref{eq:kzb2} and a Bethe-ansatz-type formula for the solution of \eqref{eq:eigv}. It would be nice to relate our formulas and theirs.

$\bullet$

The folded case would lead to the so far unnoticed two-species generalization of the elliptic
Calogero-Moser system: 
\begin{multline}
{\hat H} = -  \frac 12 {\ve}_{1} \sum_{\alpha \in [n]} {\nabla}_{w_{\alpha}}^{2} - \frac 12 {\ve}_{3} \sum_{\beta \in [m]} {\nabla}_{u_{\beta}}^{2}    + \\
+ ( {\ve}_{3} + {\ve}_{1}) \cdot  \left(  \frac{{\ve}_{3} }{{\ve}_{1}} \sum_{{\alpha}' < {\alpha}''}  \, {\wp} \left( w_{\alpha'}/w_{\alpha''} ; {\tau} \right)   + \frac{{\ve}_{1}}{{\ve}_{3}} \sum_{{\beta}' < {\beta}''}  \, {\wp} \left( u_{\beta'}/u_{\beta''} ; {\tau} \right) +  \sum_{{\alpha},  {\beta}}  \, {\wp} \left( w_{\alpha}/u_{\beta} ; {\tau} \right)
 \right) 
 \end{multline}
whose trigonometric limit was studied in \cite{sergeev}. 
\vfill\eject

\section{Appendix A. Notations}

Here we summarize our notations. 

\subsection{Roots of unity}

\beq
\ii = \sqrt{-1}\ ,
\eeq 
and
\beq
{\varpi}_{p} = {\exp}\, \frac{2\pi\ii}{p}
\label{eq:primrp}
\eeq
so that ${\ii} = {\varpi}_{4}, - {\ii} = {\varpi}_{4}^{3}$.

\subsection{Finite sets}

We denote by $[a,b]$ the set
\beq
[a,b] = \{ \, n \, | \, a \leq n \leq b, n \in {\BZ} \, \}
\eeq
For the finite set $J \subset {\BZ}$ we define $h: J \to [0, {\#}J-1 ]$ the
height function:
\beq
h_{j} = \# \, \{ \, j' \, | \, j' \in J\, , j' < j \, \}
\label{eq:hf}
\eeq

\subsection{Useful sums}

Let $(z_{i}, p_{i} ) \in {\BC}^{\times} \times {\BC}$, $i = 0,  \ldots , p$.
Let $e_{j}$, $j = 0, \ldots, p+1$, $e_{0} = 1$, be the elementary symmetric
functions of $z$'s:
\beq
e_{j} = \sum_{J \subset [ 0, p], \, {\#}J = j } \prod_{j \in J} z_{j}
\label{eq:symz}
\eeq
We have:
\beq
\prod_{j=0}^{p} ( 1 + t z_{j} ) = \sum_{J \subset [0,p]}  t^{{\#}J} \prod_{j \in J} z_{j} 
\eeq
Define:
\beq
{\CalD}^{p}_{k} = \prod_{j=0}^{p} \, \frac{1}{1 + t z_{j}} \, \sum_{J \subset [0,p]} t^{{\#}J} \prod_{j \in J} z_{j} \, \sum_{j \in J} h_{j}^{k} p_{j}
\eeq
It is easy to prove by induction, that:
\beq
\begin{aligned}
& {\CalD}^{p}_{0} = \sum_{j=0}^{p} \frac{t z_{j}}{1+ t z_{j}} p_{j} \, \qquad {\CalD}^{p}_{1} = \sum_{0 \leq i < j \leq p} \frac{t z_{i}}{1+ t z_{i}}\frac{t z_{j}}{1+ t z_{j}} p_{j} \\
& {\CalD}^{p}_{2} = {\CalD}^{p}_{1} + 2 \sum_{0 \leq i < j < k \leq p} \frac{t z_{i}}{1+ t z_{i}}\frac{t z_{j}}{1+ t z_{j}}  \frac{t z_{k}}{1+ t z_{k}}p_{k} \\
\end{aligned}
\label{eq:dspk}
\eeq
\subsection{$\bS$-functions}

The functions
\begin{equation}
{\bS}_{a,b}(x) = 1 + \frac{{\ve}_{a}{\ve}_{b}}{x ( x+ {\ve}_{a}+{\ve}_{b})}, \quad {\bS}_{+; a,b}(x) = 1 + \frac{{\ve}_{a}}{x}, \quad {\bS}_{-; a,b}(x) = 1 - \frac{{\ve}_{a}}{x+{\ve}_{a}+{\ve}_{b}},
\label{eq:uker}
\end{equation}
play a prominent r\^ole. They are related to each other, 
\beq
{\bS}_{-;a,b}(x) = {\bS}_{+;a,b}( - {\ve}_{a}-{\ve}_{b} - x) \, , 
\eeq
\beq
{\bS}_{a,b}(x) = {\bS}_{+;a,b}(x) {\bS}_{-;a,b}(x)\, .
\eeq

\section{Appendix B. Partitions and partition sums}

\subsection{Partitions}

A partition ${\lambda}$ is a finite sequence ${\lambda} = ({\lambda}_{1} \geq {\lambda}_{2} \geq \ldots \geq {\lambda}_{{\ell}({\lambda})} > 0)$ of integers, ${\lambda}_{i} \in {\BZ}_{\geq 0}$,  which we sometimes extend to an infinite non-increasing sequence of integers which stablizes at $0$, ${\lambda}_{i} = 0$, $i \gg 0$.  The number of non-zero
entries is called the {\it length} of the partition $\lambda$:
\beq
{\ell}({\lambda}) = \# \{ \, i \, \vert \, {\lambda}_{i} > 0 \, \}
\label{eq:lenlam}
\eeq
the sum of its elements is called the {\it size}
\beq
| {\lambda} | = \sum_{i=1}^{\infty} {\lambda}_{i}
\label{eq:sizlam}
\eeq
The {\it square} $\square$ is the pair $(i,j)$ of integers obeying $i,j \geq 1$. The square $\square$ belongs to $\lambda$, ${\square} \in {\lambda}$, if  
$j \leq {\lambda}_{i}$ or, equivalently if $i \leq {\lambda}_{j}^{t}$. The collection of the squares of the partition $\lambda$ is its {\it Young diagram} (the uppermost square on the left is $(1,1)$):

\bigskip
\bigskip
\centerline{\picti{5}{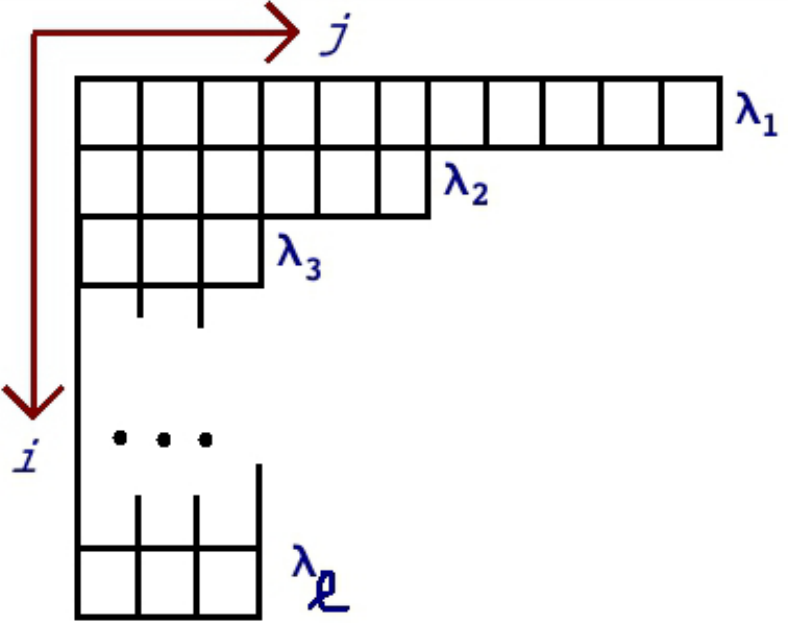}}
\bigskip
\vbox{\centerline{\bf Fig. 6}
\centerline{\sl Young diagram with the $(i,j)$-coordinates}}
\bigskip
\bigskip 

\subsubsection{The contents}

The {\it content} $c_{\square}$ of ${\square}$ is defined as:
\beq
c_{\square} = {\ve}_{1}(i-1) + {\ve}_{2} (j-1) \ .
\label{eq:cont}
\eeq
More precisely, the content \eqref{eq:cont} is defined relative to the weights $({\ve}_{1}, {\ve}_{2})$. Below, we shall use the contents defined relative to other pairs of weights as well. To avoid any confusion we either denote them by different letters, e.g.
\beq
{\sigma}_{\square} = {\ve}_{3} (i-1) + {\ve}_{4} (j-1) \ ,
\eeq
 or  explicitly specify the pair of the weights, e.g.
 \beq
 c^{ab}_{\square} = {\ve}_{a} (i-1) + {\ve}_{b} (j-1)
 \label{eq:contab}
 \eeq 
\bigskip

\centerline{\picti{9}{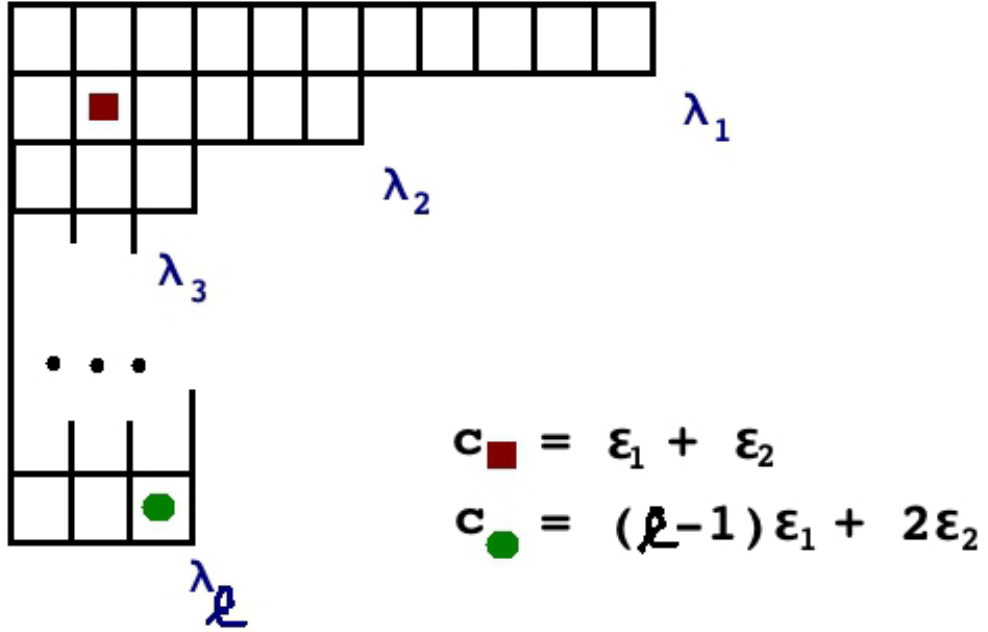}}
\bigskip
\vbox{
\centerline{\bf Fig. 7}

\centerline{\sl Examples of the contents, with ${\lambda}_{l} = 3$}} 
\bigskip
sub{Arms, legs, and hooks}

For each $\square = (i,j)  \in \lambda$ one defines the arm-length, the leg-length and the hook-length:~ $a_{\square}$, $l_{\square}$, $h_{\square}$ by:
\begin{equation}
a_{\square} = {\lambda}_{i}  - j, \qquad l_{\square} = {\lambda}_{j}^{t} - i, \qquad h_{\square} = a_{\square}+l_{\square}+1, 
\label{eq:alh}
\end{equation} 
Here is the picture for the arm and for the leg of a square $(3,4)$ in the partition $(15, 9, 9, 9, 9, 7, 4, 1, 1)$:

\bigskip
\centerline{\picti{6}{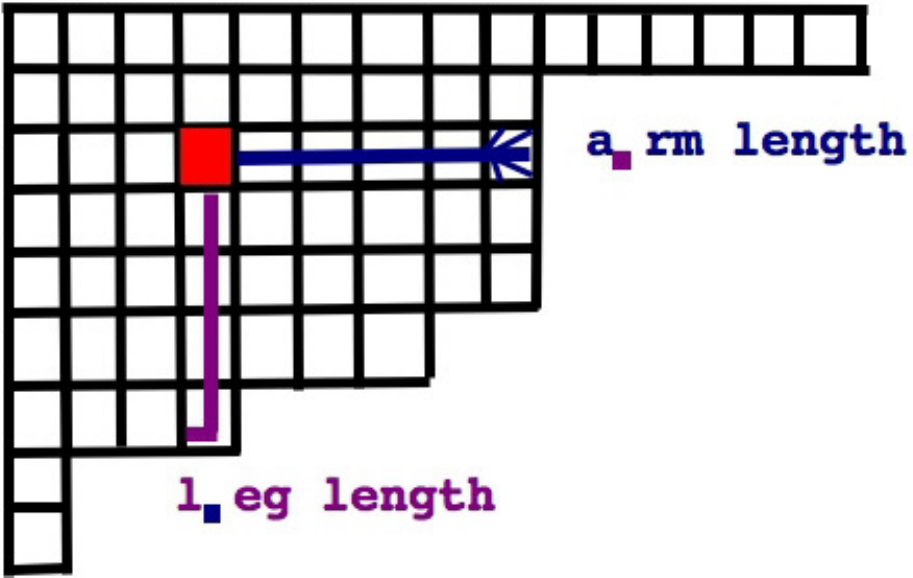}}
\bigskip
\vbox{
\centerline{\bf Fig. 8}}
\bigskip

\subsubsection{The character}

The character ${\chi}_{\lambda} (q_{1}, q_{2})$ of the partition $\lambda$ is the polynomial
in $q_1, q_2$
\beq
 {\chi}_{\lambda} (q_{1}, q_{2}) = \sum_{\square \in {\lambda}} q_{1}^{i-1}q_{2}^{j-1} 
 \label{eq:charpart}
 \eeq
 which we can also express as the sum of exponentials of contents
 using \eqref{eq:qeps} :
 \beq
 {\chi}_{\lambda} (q_{1}, q_{2}) = \sum_{\square \in {\lambda}} e^{{\beta}c_{\square}} 
 \eeq
 The character contains all information about the partition. In particular,
 \beq
 | {\lambda} | = {\chi}_{\lambda} (1,1), \quad {\ell}({\lambda}) = {\chi}_{\lambda}(1,0), \quad {\lambda}_{1} = {\chi}_{\lambda} (0,1)
 \eeq
 
\subsubsection{Dual partition}

Fix a partition $\lambda$.  The {\it dual} or {\it transposed} partition ${\lambda}^{t}$ is defined by the property:
 \beq
 {\chi}_{{\lambda}^{t}}(q_{1}, q_{2}) = {\chi}_{\lambda}(q_2, q_1)
 \eeq
 In other words,
 \beq
 {\lambda}_{j}^{t} = \# \{ \, i\,  \, | \, {\square} = (i,j) \in {\lambda} \, \Leftrightarrow \, j \leq {\lambda}_{i} \, \}
 \label{eq:dualpart}
 \eeq 
 Obviously:
 \beq
 {\ell}({\lambda}^{t})  = {\lambda}_{1}, \quad | {\lambda}^{t} | = | {\lambda} | 
 \eeq
 
 \subsubsection{Bosonic representation}
 
 A partition $\lambda$ can also be identified with a finite sequence of non-negative integers $(k_{1} k_{2} k_{3} \ldots k_{l} \ldots )$
 \beq
 {\lambda} \leftrightarrow \left(1^{k_{1}}2^{k_{2}} 3^{k_{3}} \ldots l^{k_{l}} \ldots \right) \, , 
 \label{eq:multp}
 \eeq
 where
 \beq
 k_{l} = {\lambda}^{t}_{l} - {\lambda}^{t}_{l+1} = 
 \# \{ \, i \, | \, {\lambda}_{i} = l \}
 \label{eq:multk}
 \eeq
 We have the obvious relations:
 \beq
 {\ell}({\lambda}) = \sum_{l=1}^{\infty} k_{l}, \qquad | {\lambda} | = \sum_{l=1}^{\infty} l k_{l}
 \eeq
 Since there is no restriction on the $k_{l}$'s apart from them being non-negative, this representation of partition identifies it with a state in the free boson Fock space.

 \subsubsection{Dual character} is defined by the conjugation ${\beta} \mapsto - {\beta}$ keeping ${\ve}_{1}, {\ve}_{2}, \ba$ intact:
 \beq
 {\chi}_{\lambda}^{*}(q_{1}, q_{2}) = {\chi}_{\lambda} (q_{1}^{-1}, q_{2}^{-1}), \qquad
  {\chi}_{\bla} ({\ba}, q_{1}, q_{2})^{*} =  {\chi}_{\bla} (-{\ba}, q_{1}^{-1}, q_{2}^{-1})
 \eeq
 
 \subsubsection{Colored partitions} 
 
 An $n$-colored partition ${\bla}$ is the collection $( {\ac}_{\alpha},  {\lambda}^{({\alpha})} )_{{\alpha}=1}^{n}$ of $n$ pairs consisting of a complex number and a partition. An $n$-colored square ${\hat\square}$ is a pair
 $({\alpha}, {\square})$, where ${\alpha} \in [n]$ and ${\square} \in {\lambda}^{({\alpha})}$. The content ${\bf c}_{\hat\square}$ of the $n$-colored square is the sum
 \beq
 {\bf c}_{\hat\square} = {\ac}_{\alpha} + c_{\square}
 \label{eq:ccont}
 \eeq
 It is defined relative to the weights $({\ba}, {\ept})$. We shall also encounter the contents ${\mathbf{\Sigma}}_{\hat\square}$ defined relative to the weights $({\bnu}, {\tilde\ept})$:
 \beq
 {\mathbf{\Sigma}}_{\hat\square}  = {\nu}_{\alpha} + {\sigma}_{\square}
 \label{eq:scont}
 \eeq
The character ${\chi}_{\bla} ({\ba}, q_{1}, q_{2})$ of the $n$-colored partition is the function
 \beq
  {\chi}_{\bla} ({\ba}, q_{1}, q_{2}) = \sum_{{\hat\square} \in {\bla}} e^{{\beta} {\bf c}_{\hat \square}}  = \sum_{{\alpha}=1}^{n} e^{{\beta}{\ac}_{\alpha}} {\chi}_{{\lambda}^{({\alpha})}} (q_{1}, q_{2}) 
  \label{eq:colchar}
  \eeq
We shall also need
\beq
   {\tilde\chi}_{\bla} ({\bnu}, q_{3}, q_{4}) = \sum_{{\hat\square} \in {\bla}} e^{{\beta} {\mathbf{\Sigma}}_{\hat \square}}  = \sum_{{\beta}=1}^{n} e^{{\beta}{\nu}_\beta} {\chi}_{{\lambda}^{({\beta})}} (q_{3}, q_{4}) 
  \label{eq:solchar}
  \eeq
Finally, 
\beq
{\chi}^{ab}_{\bla} = \sum_{{\alpha}=1}^{n_{ab}} e^{{\ac}_{ab, \alpha}} {\chi}_{{\lambda}_{ab}^{({\alpha})}} (q_{a}, q_{b})
\label{eq:solcharab}
\eeq

\subsection{Simple partition sums and elliptic functions}
 
\subsubsection{Generating functions of lengths and sizes}

The following generating functions are computed in an elementary fashion using the bosonic representation \eqref{eq:multp}: that of  the number of partitions
\beq
\sum_{n=1}^{\infty} p(n) {\qe}^{n} = \sum_{\lambda} {\qe}^{|{\lambda}|} = \frac{1}{{\phi}({\qe})} \, , \eeq
where
\beq 
{\phi}({\qe}) = \prod_{n=1}^{\infty} ( 1 - {\qe}^{n})\  , 
\label{eq:phif}
\eeq
and the more refined generating function
\beq
\sum_{\lambda} t^{{\ell}({\lambda})} {\qe}^{|{\lambda}|} = \frac{1}{f(t, {\qe})} \, , 
\eeq
with
\beq
f (t , {\qe} ) = \prod_{n=1}^{\infty}  ( 1 - t {\qe}^{n} ) 
\label{eq:ftq}
\eeq

\subsubsection{Elliptic functions}

Let us fix our notations for the functions $\eta$, $\theta$, $\wp$. Write
\beq
{\qe} = e^{2\pi \ii \tau}, \qquad {\Im}{\tau} > 0 
\label{eq:qetau}
\eeq
Then, define the Dedekind eta
\beq
{\eta}({\tau}) = e^{\frac{{\pi}{\ii} {\tau}}{12}} {\phi}({\qe})
\label{eq:eta}
\eeq
We shall slightly abuse the notation for the odd theta function
\beq
{\theta}_{11}(z; {\tau}) = {\ii}\, e^{\frac{{\pi}{\ii} {\tau}}{4}}\, z^{\frac 12} \ \left(  1 - z^{-1} \right) {\phi}({\qe}) f( z, {\qe}) f ( z^{-1}, {\qe}) \, , 
\label{eq:theta}
\eeq
since in most of what follows the sign ambiguity in ${\ii}\sqrt{z}$ will cancel out. 
The series expansion of \eqref{eq:theta}
\beq
{\theta}_{11}(z; {\tau}) = {\ii} \sum_{r \in {\BZ}+{\frac 12}}\, (-1)^{r - \frac 12} \, z^{r} e^{{\pi}{\ii} {\tau} r^{2}}
\label{eq:sertheta}
\eeq
implies the heat equation:
\beq
\frac{1}{{\pi}\ii} \frac{\partial}{\partial\tau}  {\theta}_{11}(z; {\tau}) = \left( z{\partial}_{z} \right)^{2} {\theta}_{11}(z; {\tau})
\label{eq:heattheta}
\eeq
Weierstrass $\wp$-function:
\beq
{\wp}(z ; {\tau}) = -\left( z{\partial}_{z} \right)^{2}\, {\log}\, {\theta}_{11}(z; {\tau})
+ \frac{1}{{\pi}{\ii}} {\partial}_{\tau}\, {\log}\, {\eta} ({\tau})
\eeq  
is normalized so as to have vanishing ${\log}(z)^{0}$ term in the expansion near $z = 1$. 

\subsubsection{Rank $p-1$ theta function}
is given by (cf. \eqref{eq:freidental})
\beq
{\Theta}_{A_{p-1}} \left({\bz} ; {\tau}\right) = {\eta}({\tau})^{p-1} \ \prod_{\alpha > \beta} \, \frac{{\theta}_{11} \left( z_{\alpha}/z_{\beta} \, ; \, {\tau} \right)}{{\eta}({\tau})}
\label{eq:thetapm}
\eeq
It obeys the \emph{heat equation} (cf. \eqref{eq:laplog} and 
\cite{Kac:1984}, \cite{Pressley:1986}, \cite{Kac:1990gs})
\beq
p\,  \frac{\partial}{d{\tau}} {\Theta}_{A_{p-1}} \left({\bz} ; {\tau}\right) \, = \ {\pi}{\ii} \, {\Delta}_{\bzv}\ {\Theta}_{A_{p-1}} \left({\bz} ; {\tau}\right)
\label{eq:heateq}
\eeq 

\subsection{More partition sums}

\noindent
{}Fix 
\[ \bar\ve = ({\ve}_{1}, {\ve}_{2}, {\ve}_{3}, {\ve}_{4}) \] 
with ${\ve}_{1} + {\ve}_{2} + {\ve}_{3} + {\ve}_{4} = 0$. 

\noindent
{}Fix a non-negative integer $p \in {\BZ}_{\geq 0}$ and 
  ${\mathfrak s}, {\omega} \in \BZ/ p{\BZ}$. 
  
{}The functions \[ \Bfn \, , \ \Qfn \, , \ \tQfn \] are  defined on the set of all partitions cf. \eqref{eq:uker} as follows:
  \beq
   {\Bfn}\,
 [{\lambda}] = 
 \prod\limits_{{\square} \in \lambda} \,  {\bS}_{+;1,2} \left({\ve}_{3}h_{\square} + {\ve} a_{\square} \right)^{{\delta}_{(p)}({\mathfrak s}\, h_{\square} + a_{\square})} {\bS}_{-;1,2} \left( {\ve}_{3}h_{\square} + {\ve} a_{\square} \right)^{{\delta}_{(p)}({\mathfrak s}\, h_{\square} + a_{\square}+1)}  \ ,  \label{eq:bomp}
 \eeq 
 and
 \beq
 \begin{aligned}
&  {\Qfn} \, [{\lambda}] \ = \ \prod\limits_{{\square} \in \lambda} \, {\qe}_{{\omega} + {\kappa}_{\square}} \ , \\
 & {\tQfn} \, [{\lambda}] \ =  \ \prod\limits_{{\square} \in \lambda} \, {\qe}_{{\omega} - {\kappa}_{\square}} \ , \label{eq:qomp}
 \end{aligned}
 \eeq
  where for ${\square} = (i,j)$
 \beq
 {\kappa}_{\square} = {\mathfrak s}(i-j) + 1-j \ ( {\rm mod} \, p) \ . 
 \label{eq:kapp}
 \eeq
Sometimes we use a shorthand notation:
\beq
{\BB}_{\omega}^{\lambda} = {\CalB}\, \left( \begin{matrix} {\ve}_{3} &   {\ve}_{4} \\
 {\ve}_{1} & {\ve}_{2} \end{matrix} \, \biggl\vert \,  \begin{matrix}   0 \\
  p \end{matrix} \, \biggr\vert\  {\bqt}\  \right)
 [{\lambda}], \qquad
 {\tilde\BB}_{\omega}^{\lambda} = {\CalB}\, \left( \begin{matrix}  {\ve}_{1} &   {\ve}_{2} \\
  {\ve}_{3} & {\ve}_{4} \end{matrix} \, \biggl\vert \,  \begin{matrix}  0 \\
  p \end{matrix} \, \biggr\vert\  {\bqt}\  \right)
 [{\lambda}] \label{eq:bbfu}
 \eeq
and
\begin{multline}
 {\BQ}_{\omega}^{\lambda} = {\CalQ}\, \left( \begin{matrix} {\omega} & 0 \\
 & p \end{matrix} \, \biggr\vert\  {\bqt}\  \right)\, [{\lambda}]  = \prod_{j}\, {\qe}_{{\omega}+1-j}^{{\lambda}_{j}^{t}} = \prod_{i} \, z_{\omega}/z_{{\omega}-{\lambda}_{i}}\ , \\ 
 {\tilde\BQ}_{\omega}^{\lambda} = {\tilde\CalQ}\, \left( \begin{matrix} {\omega} & 0 \\
 & p \end{matrix} \, \biggr\vert\  {\bqt}\  \right)\, [{\lambda}] = \prod_{j}\, {\qe}_{{\omega}+j-1}^{{\lambda}_{j}^{t}} = \prod_{i} \, z_{{\omega}+{\lambda}_{i}-1}/z_{{\omega}-1} \\
\end{multline}
Define the partition sums
\beq
{\BB}_{p,\omega} ({\bar\ve}; {\bqt})  = \sum_{\lambda} \ {\BQ}_{\omega}^{\lambda}\, {\BB}_{\omega}^{\lambda}\, , \qquad {\tilde\BB}_{p,\omega} ({\bar\ve}; {\bqt})  = \sum_{\lambda} \ {\tilde\BQ}_{\omega}^{\lambda}\, {\tilde\BB}_{\omega}^{\lambda} \ . 
\label{eq:bombtom} \eeq
When there is no confusion about what $p, {\ept}, {\mt}$ are, we use the shorter notations
 \beq
{\BB}_{\omega} \equiv {\BB}_{p,\omega} ({\ept},{\mt}, - {\mt} - {\ve}; {\bqt}), \quad {\tilde\BB}_{\omega} \equiv {\tilde\BB}_{p,\omega} ({\ept},{\mt}, - {\mt} - {\ve};  {\bqt})  \label{eq:shbb}
\eeq

\subsection{Trigonometric limit}

By the trigonometric limit we understand the limit ${\qe} \to 0$ with ${\bzv}$ kept fixed. In this limit the partition sums ${\BB}_{p,\omega} ({\bar\ve}; {\bqt}), \, {\tilde\BB}_{p,\omega} ({\bar\ve}; {\bqt})$ are elementary. Indeed, the Eq.  \eqref{eq:vecqe} implies ${\qe}_{0} = 0$.

Thus, only the partitions 
with ${\lambda}_{1} \leq \om < p$ contribute to the sum for ${\BB}_{p,\omega} ({\bar\ve}; {\bqt})$. This restriction implies that for all 
$\square \in \lambda$, $a_{\square} < {\om} \leq p-1$. Therefore only $\square = ( i , {\lambda}_{i})$ contribute to the 
${\BB}_{\om}^{\lambda}$ measure:
\begin{multline}
{\BB}_{p,\omega} ({\bar\ve}; {\bzv})^{\rm trig}  \equiv \\
 {\BB}_{p,\omega} ({\bar\ve}; {\bqt})_{\qe \to 0} = 
\sum_{{\lambda}, \, {\lambda}_{1} \leq \om} \prod_{i=1}^{{\ell}_{\lambda}} \, \frac{z_{\om}}{z_{\om - \lambda_i}} \, {\bS}_{+;1,2} ( {\ve}_{3} h_{\square} )  =
\sum_{{\lambda}, \, {\lambda}_{1} \leq \om} \, \prod_{l=1}^{{\lambda}_{1}}
\prod_{h=1}^{{\lambda}_{l}^{t} - {\lambda}_{l+1}^{t}} \left( \frac{z_{\omega}}{z_{{\omega}-l}}\,  {\bS}_{+;1,2} ({\ve}_{3} h) \right) =  \\
\sum_{{\lambda}, \, {\lambda}_{1} \leq \om} \, \prod_{l=1}^{{\lambda}_{1}}
\frac{\left( {\lambda}_{l}^{t} - {\lambda}_{l+1}^{t}  + \frac{\ve_1}{\ve_3} \right) !}{\left( {\lambda}_{l}^{t} - {\lambda}_{l+1}^{t} \right) ! \left(  \frac{\ve_1}{\ve_3} \right) !} 
\, \left( \frac{z_{\omega}}{z_{\omega - l}} \right)^{{\lambda}_{l}^{t} - {\lambda}_{l+1}^{t}}\ = \\
\sum_{k_{1}, k_{2}, \ldots, k_{\omega - 1} \geq 0}
\prod_{l=1}^{{\omega}-1} \, \left( \begin{matrix}
 k_{l}  + \frac{\ve_1}{\ve_3} \\ k_{l} \end{matrix} \right) \, \left( \frac{z_{\omega}}{z_{\omega - l}} \right)^{k_{l}} \ =\ \prod_{l=1}^{{\om}-1} \left( 1 - \frac{z_{\om}}{z_{{\om}-l}} \right)^{-1-\frac{{\ve}_{1}}{{\ve}_{3}}}\, ,  
\label{eq:bomtrig}
\end{multline} 
where we used the bosonic representation of the partition, cf. \eqref{eq:multp}:
\beq
{\lambda} \leftrightarrow \left( 1^{k_{1}} 2^{k_{2}} \ldots ({\omega}-1)^{k_{\om - 1}} \right)
\label{eq:multlk}
\eeq
Similarly, only the partitions $\lambda$ with $\lambda_{1} \leq p - {\om}$ contribute to ${\tilde\BB}_{p,\omega} ({\bar\ve}; {\bqt})$ in the trigonometric limit, thus we can use the bosonic representation \eqref{eq:multp} to perform the sum explicitly,  
\beq
{\tilde\BB}_{p,\omega} ({\bar\ve}; {\bzv})^{\rm trig}  \equiv
 {\tilde\BB}_{p,\omega} ({\bar\ve}; {\bqt})_{\qe \to 0}  =\ \prod_{l=1}^{p-{\om}} \left( 1 - \frac{z_{\om + l - 1}}{z_{{\om}-1}} \right)^{-1-\frac{{\ve}_{3}}{{\ve}_{1}}}\, ,  
\label{eq:btomtrig}
\eeq 

\subsection{The roots-weights coordinates and moduli spaces}

Let $p$ be a positive integer, and 
\beq
{\bqt} = ( {\qe}_{0} , {\qe}_{1}, \ldots , {\qe}_{p-1} ) = ( {\qe}_{\omega} )_{{\omega} = 0}^{p-1}
\in \left( {\BC}^{\times}\right)^{p}
\eeq  a collection of non-zero complex numbers. 
We shall be dealing with generating functions of the form:
\beq
Z_{A_{p-1}} ({\bqt}) = \sum_{{\bkt} \in {\BZ}_{+}^{p}} {\bqt}^{\bkt} \ Z_{A_{p-1}} [{\bkt}]
\label{eq:zapi}
\eeq
where $Z_{A_{p-1}} [ k_{0}, \ldots , k_{p-1} ]$ may represent a contribution of $p$ ``fractional'' instantons of the charges $k_{0}$, $k_{1}$, ... , $k_{p-1}$, or a contribution of $p$ instantons belonging to $p$ different gauge groups with an $A$-type quiver interaction between them.  The geometric or, physically, weak coupling domain of definition of the generating function \eqref{eq:zapi} is the polydisk ${\CalD}^{\rm weak}$: $| {\qe}_{\omega} | < 1$, ${\omega}  = 0 , \ldots , p-1$. 

Surprisingly enough, we shall see that an analytic continuation of \eqref{eq:zapi} outside
${\CalD}^{\rm weak}$ is often possible.  
In the fractional case we shall see an emergence of the Weyl symmetry $S_{p}$ of the $A_{p-1}$-type, even though the original problem didn't have this symmetry in a manifest way.  In the $A$-type quiver case we shall see the emergence of the modular group of the ${\CalM}_{0,p+2}$ moduli space of $p+2$-punctured genus zero curves. 

The variables ${\bqt}$ are
not, however, convenient for exhibiting these symmetries. 

\subsubsection{Fractional case}

Let us extend $\bqt$ to a $p$-periodic function ${\BZ} \to {\BC}^{\times}$: 
\beq
{\qe}_{{\omega} + p} = {\qe}_{\omega} \ .
\eeq 
Define $p+1$ variables 
\beq
{\bzv} = ( z_{0}, z_{1}, \ldots, z_{p-1}) = ( z_{\om} )_{{\om} = 0}^{p-1} \in \left( {\BC}^{\times}\right)^{p} \ ,
\eeq
and ${\qe} \in {\BC}^{\times}$ by
\beq
{\qe}_{\om} = z_{\om}/z_{{\om}-1}, \qquad {\om} = 1, \ldots, p-1, \qquad
{\qe}_{0} = {\qe}\, z_{0}/z_{p-1} 
\label{eq:vecqe}
\eeq
Thus, 
\beq
{\qe} = {\qe}_{0}{\qe}_{1} \ldots {\qe}_{p-1} = {\qe}_{[p]} = {\qe}_{[0,p-1]}\ , 
\eeq
while $z_{\om}$'s are defined up to an overall rescaling:
\beq\begin{aligned}
& z_{1} = z_{0} \, {\qe}_{1}  \, , \\
& z_{2} = z_{0} \, {\qe}_{1} {\qe}_{2} \, ,  \\
& \ldots \, , \\
& z_{p-1} = z_{0} \, {\qe}_{1}{\qe}_{2} \ldots {\qe}_{p-1} \, , \\
\end{aligned}
\label{eq:zcor}
\eeq
It is natural to extend $\bz$ to the quasi-periodic function on $\BZ$:
\beq
z_{{\om}+p}\, = \, {\qe} \ z_{\om}, \qquad {\om} \in {\BZ}
\label{eq:zper}
\eeq
The cyclic permutation of the variables ${\qe}_{\omega}$, i.e.
\beq
{\qe}_{\om} \mapsto {\qe}_{{\om}+a}, \qquad a = 0, 1, \ldots, p-1
\eeq
acts on $z_{\om}$'s via:
\beq
\begin{aligned}
& z_{\om}\ \mapsto \ z_{{\om}+a}, \qquad {\om} = 0, \ldots, p-1-a \\
& z_{\om} \ \mapsto \ {\qe} \, z_{{\om} +a - p}, \qquad {\om} = p-a, \ldots , p-1
\end{aligned}
\label{eq:shifts}
\eeq
It is on the variables ${\bzv}$ that $S_{p}$ will act by permutations in the fractional case. Moreover, when the variables $\bzv$ are extended as in \eqref{eq:zper}, one gets an action of the affine Weyl group $S_{p} \sdtimes {\BZ}^{p}$ by arbitrary finite permutations of $(z_{\omega})_{{\omega} \in {\BZ}}$. In this way the ${\bzv}$ defined up to the overall rescaling and these extended permutations (keeping $\qe$ fixed)  parametrize
the coarse moduli space  $Bun_{E} (PGL_{p})$ of holomorphic $PGL_{p}$ bundles on the elliptic curve $E = {\BC}^{\times}/{\qe}^{\BZ}$. 

\subsubsection{Linear quiver case}

In the $A_{p-1}$-type quiver case, define $\bzv \in \left( {\BC}^{\times} \right)^{p}$ via
\beq\begin{aligned}
& z_{1} = z_{0} \, {\qe}_{1}  \, , \\
& z_{2} = z_{0} \, {\qe}_{1} {\qe}_{2} \, ,  \\
& \ldots \, , \\
& z_{p-1} = z_{0} \, {\qe}_{1}{\qe}_{2} \ldots {\qe}_{p-1} \, , \\
\end{aligned}
\label{eq:zcorq}
\eeq
We shall 
think of $\bzv$ as of the coordinates
on the cell $0 < |z_{p-1}| < |z_{p-2} | < \ldots < |z_{1}| < |z_{0}| < \infty$ in the moduli space ${\CalM}_{0,p+2}$, which is the space of $p+2$-tuples of distinct points  $z_{-1}, z_{0}, \ldots , z_{p-1}, z_{p}$ on ${\BC\BP}^{1}$ modulo the overall $PGL(2)$ action. By fixing the first point $z_{-1}$ to be $z_{-1} = \infty$ and the last point $z_{p}$ to $z_{p} = 0$ one ends up with the remaining ${\BC}^{\times}$ symmetry of the overall rescaling of $\bzv$, which can be fixed, e.g. by setting $z_{0} = 1$. 

\subsubsection{${\hat A}_{p-1}$-quiver case}

The formulas \eqref{eq:zcor}, without any permutation symmetry of the $z_{i}$'s, can be interpreted as parameterizing the open cell in the moduli space ${\CalM}_{1,p}$, of $p$-punctured elliptic curves. 
The modulus ${\qe} = {\exp} \, 2\pi \ii \tau$ defines the elliptic curve
${\CalE} = {\BC} / 2\pi \ii \left( {\BZ}\oplus {\tau}{\BZ} \right) \approx {\BC}^{\times} / {\qe}^{\BZ}$, while $z_{0}, z_{1}, \ldots, z_{p-1}$ determine the location of punctures. This identification is further supported by the modular properties of the partition functions we study in this paper, they transform nicely (albeit in a complex fashion)
under the action of $SL_{2}({\BZ})$:
\beq
{\tau} \mapsto \frac{a \tau  + b}{c \tau +d}\, , \qquad
z_{i} \mapsto {\exp} \ \frac{{\log} (z_{i} )}{c \tau + d} \label{eq:modular}
\eeq
with $a, b, c, d \in {\BZ}$ obeying
\[ ad - bc = 1 \ . \]

\subsubsection{Differential operators}
For ${\omega} = 0, \ldots, p-1$ let 
\beq
{\nabla}_{\omega}^{\qe} =  {\qe}_{\omega} \frac{\partial}{{\partial}{\qe}_{\omega}}\ ,
\label{eq:qder}
\eeq 
\beq
{\nabla}_{\omega}^{z} =  z_{\omega} \frac{\partial}{{\partial}z_{\omega}} \ , 
\label{eq:zder}
\eeq
Let $\Delta_{\bzv}$ denote the Laplacian in ${\log}(z_{\omega})$'s:
\beq
{\Delta}_{\bzv} = {\nabla}^{\bzv} \cdot {\nabla}^{\bzv}  \equiv \sum_{{\om}=0}^{p-1} \left( {\nabla}^{z}_{\om} \right)^{2} 
\label{eq:laplog}
\eeq
Let
\beq
{\Nq} = {\qe} \frac{\partial}{\partial \qe} \ , 
\label{eq:deder}
\eeq
when acting on the function of $({\bzv}, {\qe})$, i.e. we keep $\bzv$ fixed when differentiating with respect to $\qe$. The definition \eqref{eq:vecqe} implies ${\Nq} = {\nabla}_{0}^{\qe}$ and:
\beq 
{\nabla}_{\omega}^{z} = {\nabla}_{\omega}^{\qe} - {\nabla}_{\omega +1}^{\qe}\, , \label{eq:derzq}
\eeq
for ${\omega} = 0, \ldots , p-1$.
Let 
\beq
{\Dq} = \sum_{{\omega}=0}^{p-1} {\nabla}^{\qe}_{\omega}
\label{eq:kzd}
\eeq
In terms of $\nabla^z$, $\Nq$ it reads:
\beq
{\Dq} = p\, {\Nq} + {\rv}\cdot {\nabla}^{\bzv} \, , 
\label{eq:kzdfrz}  
\eeq
where
\beq
{\rv}\cdot {\nabla}^{\bzv} \equiv 
\sum_{{\omega}=0}^{p-1}{\rho}_{\omega} {\nabla}_{\omega}^{z}\, , 
\eeq
and 

\subsubsection{The rho-vector}

Define
\beq
{\rv} = ( {\rho}_{0}, {\rho}_{1}, \ldots , {\rho}_{p-1} ) \in \left( \frac12 \, {\BZ} \right)^{p}
\label{eq:rhov}
\eeq
where
\beq
{\rho}_{\omega}  = {\omega} - \frac{p-1}{2}, 
\label{eq:rhovii}
\eeq
\beq
{\rv}^{2} = \frac{p(p^{2}-1)}{12} \label{eq:freidental}
\eeq
We also use the notation
\beq
{\bzv}^{\rv} \  =\   \prod_{{\omega}={0}}^{p-1} \, z_{\omega}^{\rho_{\omega}}
\label{eq:zpr}
\eeq

\subsection{Differential identities}

In the main body of the paper we shall prove
the identity
\beq
{\ve}_{1} {\nabla}_{\omega'}^z {\log}({\tilde\BB}_{p,\omega''} ({\bar\ve}; {\bqt}) ) + 
{\ve}_{3} {\nabla}_{\omega''-1}^z {\log} ({\BB}_{p,\omega'} ({\bar\ve}; {\bqt})) = 0 , \qquad
{\omega}'' - {\omega}' \neq 1 \, {\rm mod}\ p
\label{eq:diffrel1}
\eeq
which implies (by summing over $\omega'$, keeping ${\omega}'' = {\omega}$ or by summing over $\omega''$, keeping ${\omega}' = {\omega}-1$, respectively)
\beq
 {\nabla}_{{\om}-1}^z \left( {\ve}_1 {\log}({\tilde\BB}_{\omega})+{\mt}\, {\log}({\BB}_{\omega-1}) \right) = {\mt} \, {\nabla}_{\omega-1}^{z}  {\log} ({\BB}) = {\ve}_{1} {\nabla}_{\omega - 1}^z {\log} ({\tilde\BB})
 \label{eq:bfrobom}
 \eeq
 where we
 defined  ${\BB}\equiv {\BB}_{p} ({\ept}, {\mt}; {\bqt}), \ {\tilde\BB}\equiv {\tilde\BB}_{p} ({\ept}, {\mt}; {\bqt})$ via: 
\beq
\begin{aligned}
& \qquad\qquad {\BB}_{p} ({\ept}, {\mt}; {\bqt})
\equiv \prod_{\omega = 0}^{p-1} \ {\BB}_{p,\omega} ({\ept}, {\mt}, - {\mt} - {\ve}; {\bqt})   \,  , \\
& \qquad\qquad {\tilde\BB}_{p} ({\ept}, {\mt}; {\bqt})
\equiv \prod_{\omega = 0}^{p-1} \ {\tilde\BB}_{p,\omega} ({\ept}, {\mt}, - {\mt} - {\ve}; {\bqt})  \\
\label{eq:bbbt}
\end{aligned}\eeq

\subsection{$\Delta$-functions}

Let us define the functions ${\BDe} = {\BDe}_{p} ({\ept}, {\mt}; {\bqt}),\  {\tilde\BDe} = {\tilde\BDe}_{p} ({\ept}, {\mt}; {\bqt})$ by:
\beq
{\log} ({\BDe} )
 = - \frac{\mt}{{\mt}+{\ve}_{1}} \, {\log}({\BB})  \ , 
  \
  {\log} ({\tilde\BDe})
 = - \frac{\ve_1}{{\mt}+{\ve}_{1}} \, {\log}({\tilde\BB}) \, 
 \label{eq:bbbtddt}
 \eeq
 which are equal up to a $\qe$-(and of course $\ve$-)dependent factor thanks to \eqref{eq:bfrobom}, 
 \beq
{\BDe}_{p} ({\ept}, {\mt}; {\bqt}) \ = \ {\tilde\BDe}_{p} ({\ept}, {\mt}; {\bqt}) \ {\exp}\, {\chi} ({\qe}; {\bar\ve}) \ .
 \label{eq:dbbbt}
 \eeq 
We shall also prove: 
\beq
{\nabla}_{\alpha}^{z}{\nabla}_{\beta}^{z}\ {\log}{\BDe} = f_{\alpha\beta} (z_{\alpha}/z_{\beta} ; {\qe})
\label{eq:weirid}
\eeq
for all ${\alpha} \neq {\beta} \in {\BZ}_{p}$, and, finally, the heat equations,
\beq
 \left(  p {\Nq} - p\frac{\ve_2}{{\mt}+{\ve}_1} {\Nq}{\chi} + {\rv} \cdot {\nabla}^{\bzv} - \frac 12 {\Delta}_{\bzv} \right) {\BDe} = 0 
 \label{eq:heateqd}
 \eeq
 from which we derive
 \beq
\mathboxit{\begin{aligned}
& {\Delta} \ =\  {\bzv}^{\rv}\, {\qe}^{-\frac{{\rv}^{2}}{2p}}\, e^{\frac{{\ve}_{2}}{{\mt}+{\ve}_{1}} {\chi}({\qe})} \, {\Theta}_{A_{p-1}} ({\bzv}; {\tau})  \\
& {\tilde\Delta} \ =\ {\bzv}^{\rv}\, {\qe}^{-\frac{{\rv}^{2}}{2p}} \, e^{\frac{{\mt}+{\ve}}{{\mt}+{\ve}_{1}} {\chi}({\qe})} \, {\Theta}_{A_{p-1}} ({\bzv}; {\tau})  \\
\end{aligned}}
\label{eq:detdeth}
\eeq

\subsection{Generalization of  Macdonald Identities}

We conjecture that the stronger statements hold (see below for the list of checks we performed):
\beq
\mathboxit{
{\BB}_{p,\omega} ({\bar\ve}; {\bqt}) =   \left( {\phi}({\qe})^{\frac{{\ve}_{1}}{p {\ve}_{4}} }\cdot {\Theta}_{\omega}({\bzv}; {\qe}) \right)^{-1 - \frac{{\ve}_{1}}{{\ve}_{3}}}\, , \qquad
{\tilde\BB}_{p,\omega} ({\bar\ve}; {\bqt}) =   \left( {\phi}({\qe})^{\frac{{\ve}_{3}}{p {\ve}_{2}} }\cdot {\tilde\Theta}_{\omega}({\bzv}; {\qe}) \right)^{-1 - \frac{{\ve}_{3}}{{\ve}_{1}}}\,
 }
\label{eq:ellid}
\eeq
(these two formulas follow one from another, by exchanging $({\ve}_{1}, {\ve}_{2}) \leftrightarrow ({\ve}_{3}, {\ve}_{4})$)
where, cf. \eqref{eq:zcor}: 
\beq
\begin{aligned}
& \qquad {\Theta}_{\omega}({\bzv}; {\qe}) = \prod_{{\alpha} = 0}^{{\omega}-1} \left( 1 - z_{\omega}/z_{\alpha} \right) \
 \prod_{{\alpha}=0}^{p-1}  \, f\left( z_{\omega}/z_{\alpha} , {\qe} \right) = \\ 
 & \\
& \qquad\qquad\qquad = \prod_{{\ell}=0}^{\infty} \left( 1 - \prod_{j=1}^{\ell} {\qe}_{{\omega}+1-j} \right) = \prod_{{\ell}=1}^{\infty} \left( 1 - z_{\omega}/z_{{\omega} - {\ell}} \right)\,  , 
\qquad \\
& \\
& \qquad {\tilde\Theta}_{\omega}({\bzv}; {\qe}) = \prod_{{\alpha} = {\omega}}^{p-1} \left( 1 - z_{\alpha}/z_{\omega - 1} \right) 
\  \prod_{{\alpha}=0}^{p-1}  \, f \left(  z_{\alpha}/z_{{\omega}-1} , {\qe} \right) = \\
& \\
& \qquad\qquad\qquad =  \prod_{{\ell}=0}^{\infty} \left( 1 - \prod_{j=1}^{\ell} {\qe}_{{\omega}+j-1} \right) = 
\prod_{{\ell}=1}^{\infty} \left( 1 - z_{{\omega}+{\ell}-1}/z_{{\omega} - 1} \right) \\
\label{eq:thetom}
\end{aligned}
\eeq
are the ``half-theta''-functions, related to each other and the $A_{p-1}$-theta function
\eqref{eq:thetapm} via
\beq
\prod_{\om \in {\BZ}_{p}} {\tilde\Theta}_{\omega}({\bzv}; {\qe}) \sim
\prod_{\om \in {\BZ}_{p}} {\Theta}_{\omega}({\bzv}; {\qe})  \sim {\phi}({\qe})\ 
{\bzv}^{\rv} {\qe}^{-\frac{{\rv}^2}{2p}} \ {\Theta}_{A_{p-1}} ({\bzv}; {\tau})
\label{eq:prdthetas}
\eeq
The symbol ``$\sim$'' in \eqref{eq:prdthetas} stands to mean that we are omitting a constant factor
like ${\ii}^{p(p-1)/2}$, which drops out of the formulas we use in the paper anyway. 

The conjecture \eqref{eq:ellid}, \eqref{eq:thetom} implies (cf. \eqref{eq:thetapm})
\beq
\mathboxit{{\chi}({\qe}; {\bar\ve}) = \left( \frac{\ve_3}{\ve_2} - \frac{\ve_1}{\ve_4} \right) \, {\log}\, ({\phi}({\qe})) }
\label{eq:chiconj}
\eeq

\subsubsection{Checks of the conjecture}

Let us make several comments about \eqref{eq:ellid}, \eqref{eq:thetom}. 
First, when ${\ve}_{3} \to \infty$, or when ${\ve}_{1} \to 0$, the left hand side of \eqref{eq:ellid} becomes:
\beq
\sum_{\lambda} \prod_{{\square} \in {\lambda}} {\qe}_{{\omega} + 1 - j\, {\rm mod}\, p} = \sum_{\lambda} \prod_{i=1}^{\infty} \frac{z_{\omega}}{z_{{\omega}-{\lambda}_{i}}} =  \prod_{{\ell}=1}^{\infty} \left( 1 - \frac{z_{\omega}}{z_{{\omega}-{\ell}}} \right)^{-1} = {\Theta}_{\omega}({\bzv}; {\qe})^{-1}
\eeq
by \eqref{eq:thetom}. Thus, \eqref{eq:ellid} holds in this limit. 

\noindent
\hbox{\vbox{\hbox{Second,
when ${\ve}_{1} + {\ve}_{3} \to 0$,}\hbox{the right hand side of \eqref{eq:ellid} approaches $1$.} \hbox{The left hand side does the same,}
\hbox{since any $\lambda \neq \emptyset$ has}
\hbox{at least one square $\square$ with $a_{\square} = l_{\square} = h_{\square} - 1 = 0$.}\hbox{Such a square contributes}\hbox{the factor  $1 + {\ve}_{1}/{\ve}_3$ to \eqref{eq:bomp},}
\hbox{which vanishes when ${\ve}_{3} = - {\ve}_{1}$.}}\vbox{\picti{3}{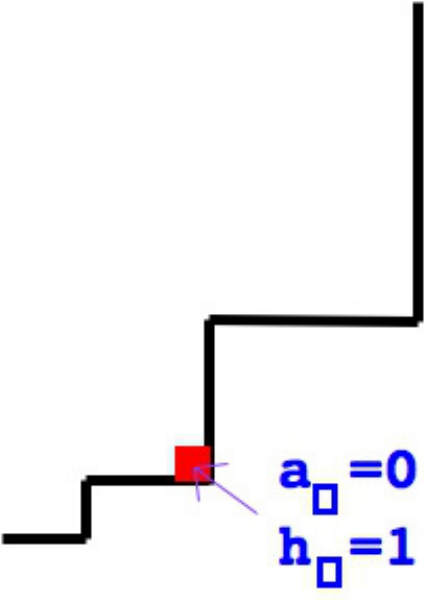}}} 

\noindent Thus, \eqref{eq:ellid} holds in this limit too. Moreover, we can compute the limit, as ${\ve}_{3} \to - {\ve}_{1}$, 
\begin{multline}
\frac{{\ve}_{1} \, {\log}({\tilde\BB}_{\om})}{{\ve}_{1}+ {\ve}_{3}} \longrightarrow \sum_{{\lambda}  = w \times h}
\, \left( \frac{z_{{\om} + w - 1}}{z_{{\om}-1}} \right)^{h} \times
\prod_{l=2}^{h} \left( 1 - \frac{1}{l} \right)  \times \\   \times \prod_{l=1}^{h} \left( 
\prod_{k=1}^{[w/p]_{-}} \left( 1 - \frac{1}{l - \frac{p{\ve}_{2}}{{\ve}_{1}} k } \right) \prod_{k=1}^{[w/p]} \left( 1 + \frac{1}{l - 1 - \frac{p {\ve}_{2}}{\ve_1} k } \right) \right)   =
\\
\qquad = \sum_{h, w =1}^{\infty} \frac{1}{h} 
\left( \frac{z_{{\om} + w - 1}}{z_{{\om}-1}} \right)^{h} \times 
\left\lbrace\, \begin{matrix}
 1 \, , &\ w \neq 0 \, ( {\rm mod}\, p ) \\
 1 -  \frac{h {\ve}_{1}}{w {\ve}_{2}}\, , & \ w \equiv 0 \, ( {\rm mod}\, p ) \end{matrix}  \right. \\
= - {\log} \, {\tilde\Theta}_{\omega}({\bzv}; {\qe})   + \frac{\ve_{1}}{p \ve_{2}} {\log} \left( {\phi}({\qe}) \right)  
\label{eq:rectangle}
\end{multline}
where we used the fact that the only partitions $\lambda$ which contribute to the limit are the ones with exactly one square $\square$ with $a_{\square} = l_{\square} = h_{\square} - 1 = 0$, i.e. exactly one south-east corner. The Young diagrams of such partitions are the $w \times h$ rectangles, with ${\ell}_{\lambda} = h$, 
${\ell}_{\lambda^t} = w$, $| {\lambda} | = w h$. In the formula  \eqref{eq:rectangle} we denote by $[x]_{-}$ the maximal integer which is strictly less then $x$, i.e. $[x]_{-} < x$ for all $x$, and by $[x]$ the usual integer part of $x$, i.e. $[x] \leq x$, and $[x] = x$, if $x \in \BZ$. 

{}Finally, when $p \to \infty$, the only squares with nontrivial contribution to \eqref{eq:bomp} with ${\mathfrak s} = 0$ are those with $a_{\square} = 0$. Now, use the bosonic representation \eqref{eq:multp}. The rows with $\lambda_i = l$ with ${\lambda}_{l+1}^{t} \leq i \leq {\lambda}_{l}^{t}$ contribute 
\beq
\prod_{h=1}^{{\lambda}_{l}^{t} - {\lambda}_{l+1}^{t}} \left( \frac{z_{\omega}}{z_{{\omega}-l}}\,  {\bS}_{+;1,2} ({\ve}_{3} h) \right) = 
\frac{\left( {\lambda}_{l}^{t} - {\lambda}_{l+1}^{t}  + \frac{\ve_1}{\ve_3} \right) !}{\left( {\lambda}_{l}^{t} - {\lambda}_{l+1}^{t} \right) ! \left(  \frac{\ve_1}{\ve_3} \right) !} 
\, \left( \frac{z_{\omega}}{z_{\omega - l}} \right)^{{\lambda}_{l}^{t} - {\lambda}_{l+1}^{t}}\ = \  \left( \begin{matrix}
 k_{l}  + \frac{\ve_1}{\ve_3} \\ k_{l} \end{matrix} \right) \, \left( \frac{z_{\omega}}{z_{\omega - l}} \right)^{k_{l}}
\label{eq:bsh}
\eeq
to  ${\BB}_{\omega}^{\lambda} {\BQ}_{\omega}^{\lambda}$. Summing over all $\lambda$'s is equivalent to summing over all $k_{1}, k_{2}, \ldots$ independently, from $0$ to $\infty$, giving rise to:
\beq
\prod_{l=1}^{\infty} \left( 1 - z_{\omega}/z_{{\omega}-l} \right)^{-1 - \frac{\ve_1}{\ve_3}}
\eeq
which is indeed the $p \to \infty$ limit of the right hand side of the Eq. \eqref{eq:ellid}.

\section{\textbf{Acknowledgements}.}\ Research was partly supported
by the National Science Foundation under grant
no.~NSF-PHY/1404446. Any opinions, findings, and conclusions or
recommendations expressed in this material are those of the authors
and do not necessarily reflect the views of the National Science
Foundation. 

The author is grateful to A.~Okounkov, V.~Pestun, F.~Smirnov and S.~Shatashvili for numerous useful discussions, and to A.~Tsymbaliuk for collaboration on \cite{NT}.

This paper concludes a mini-series of papers \cite{N3,N4,N5,N6} dedicated to the BPS/CFT-correspondence. Since we began presenting our findings at various seminars and conferences, e.g. \cite{N1, NBd, NeCM} a lot of interesting papers appeared, e.g. \cite{G2, GK, Kanno:2013aha, Kimura:2015rgi, Bourgine:2015szm, Bourgine:2016vsq,Poghossian:2016rzb, Poghosyan:2016mkh} where some of our results have been derived as well. Also the papers 
\cite{Jeong:2017pai, Jeong:2017mfh} contain some of the further developments of these ideas.

\vfill\eject

\end{document}